\documentclass{aa}
\usepackage{lscape}
\usepackage{graphicx}
\usepackage{txfonts}
\usepackage[english]{babel}  
\usepackage{graphicx}  
\usepackage[utf8]{inputenc}  
\usepackage{microtype}  
\usepackage{booktabs}  
\usepackage{rotating}  
\usepackage{lscape}
\usepackage[squaren]{SIunits}
\usepackage{natbib}
\usepackage{color}
\usepackage{amsmath}
\usepackage{amssymb}
\usepackage{soul} 
\usepackage{hyperref}

\newcommand{\water}{H$_2$O}
\newcommand{\micron}{$\mu$m}

\newcommand{\Lsun}{L$_{\sun}$}
\newcommand{\Myr}{M$_{\sun}$\,yr$^{-1}$}
\newcommand{\mdot}{$\dot{M}$}

\newcommand{\Lup}{$L_{\rm H_2O}^{\rm up}$}

\newcommand{\tB}[1]{\textcolor[rgb]{0.0,0.0,1.0}{#1}}

\newcommand{\kms}{km\,s$^{-1}$}

\newcommand{\Lsol}{L$_{\odot}$}

\newcommand{\gsim}{\;\lower.6ex\hbox{$\sim$}\kern-7.75pt\raise.65ex\hbox{$>$}\;}
\newcommand{\lsim}{\;\lower.6ex\hbox{$\sim$}\kern-7.75pt\raise.65ex\hbox{$<$}\;}

\begin{document} 

\title{Water vapour masers in long-period variable stars}
\subtitle{IV. Mira variables 
\object{R\,Cas}, \object{o\,Cet} and \object{R\,Leo}}

\author {J.~Brand \inst{1,2}
        \and D.~Engels \inst{3}
        \and A. Zanichelli \inst{1}
        \and S. Etoka \inst{4}
        }

\institute{INAF - Istituto di Radioastronomia, Via P. Gobetti 101,
           I--40129 Bologna, Italy 
      \and Italian ALMA Regional Centre, INAF-IRA, Bologna 
      \and Hamburger Sternwarte, Universit\"{a}t Hamburg, Gojenbergsweg 112,
           D--21029 Hamburg, Germany
     \and Jodrell Bank Center for Astrophysics, Univ. of Manchester, Manchester, UK
           }

\date{Received date; accepted date}

\abstract 
{We carry out decades-long single-dish monitoring of the variation in water-maser emission associated with the circumstellar envelopes (CSEs) of different types of evolved stars.}
{We follow the variation in the maser emission over long time intervals (multiple optical periods) to understand the structure and kinematics of the stellar wind that drives the maser clouds away from the star. We also determine how this may depend on the stellar properties.
}
{We carried out monitoring campaigns with single-dish telescopes of water-maser emission at 22~GHz in the CSEs of four stars: o\,Cet, R\,Leo, $\chi$\,Cyg, and R\,Cas. The observations took place with some interruptions between 1987 and 2023. The exact time interval differed from one star to the next, but no star was monitored for fewer than 15~years. 
}
{The variability in integrated flux in the masers in R\,Cas and o\,Cet followed the variability in the optical with the same period, but with a lag of about one-third in phase. R\,Leo was too often below our sensitivity threshold for us to determine a radio period. Remarkably, no maser at all was detected in $\chi$\,Cyg. The variability in the masers in R\,Cas has a distinctive pattern. The total flux, modulated by the pulsations of the star, gradually increases to a maximum, which is followed by a similar decrease. This takes about 20~years. The pattern is repeated after an interval of quiescence of several years. Our observations have covered about one and a half cycle of this pattern so far. During its decline from a maximum, the variation in the flux resembles a damped harmonic oscillator. There are two dominant emission components that move almost tangentially on either side of the star with respect to the observer. The redshifted component likely originates from a single cloud and seems to be falling back towards the star. The blue component, moving in the CSE hemisphere nearest to us, has no drift in the line-of-sight velocity and appears to originate in a time series of short-living clouds with line-of-sight velocities within $\sim 1$~\kms\ of each other. No systematic velocity drifts are found in R\,Leo and o\,Cet. A few bursts of emission were detected at infrequent times in R\,Cas and R\,Leo that lasted about a year and caused an increase in the flux density by $1-2$ orders of magnitude. The velocity range of the maser emission is $\lsim$ 10~\kms, which is narrower than the majority of the Miras and semi-regular variables we studied so far. In particular, in the stars with a low bolometric luminosity, o\,Cet and R\,Leo, only the brightest maser components from a limited part of the CSE are visible. 
}
{The existence of a zone in the CSE with favourable conditions for maser excitation is confirmed most clearly in the case of R\,Cas through the unique pattern of its maser variability. The bolometric luminosity of a star and the velocity range of its water-maser emission, that is, the number of emission components in the spectra, are clearly correlated. Most maser components in the CSEs we studied originated in clouds that move almost perpendicular to the line of sight. The redshifted emission in R\,Cas is consistent with an origin in a single cloud that lived for at least about eight years.
}


\maketitle


\section{\label{intro} Introduction}
This is the fourth in a series of papers in which we present the results of decade-long single-dish monitoring of water-masers in the circumstellar envelopes (CSEs) of semi-regular variables (SRVs), Mira variables, OH/IR stars, and red supergiants (RSGs). The long-term monitoring was motivated by the need to capture the temporal variability and evolution of maser features over multiple pulsation cycles. Extended campaigns like this allow us to trace the morphology and kinematics of the emission and enable us to identify patterns, periodicities, and anomalies in maser behaviour, based on which, we unravel the evolution of structures in the underlying stellar wind with time. Our Medicina/Effelsberg programme started in 1987 using the 32\,m Medicina and the 100\,m Effelsberg antennas and ran until 2011. Some additional observations were made in 2015. A new programme on a smaller sample was started at Medicina in 2018, and this Medicina Long Project (MLP) will continue until 2027 at least.

In the first three papers, we presented the results for the SRVs RX\,Boo and SV\,Peg (\citealt{winnberg08}; hereafter, Paper I) and R\,Crt and RT\,Vir (\citealt{brand20}; hereafter, Paper II). In the third paper (\citealt{winnberg24}; hereafter, Paper III), we presented U\,Her and RR\,Aql as representatives of the class of Mira variables. In all stars, water-maser variability manifests itself on different timescales: The prevalent variation is periodic in the Miras and follows the optical variability of the stars with a lag of two to three months. Miras and SRVs show irregular fluctuations with durations of a few months and systematic variations on timescales of a decade or longer. 
The lifetimes of single maser clouds (corresponding to individual peaks in the spectra) were found to be about a few ($\lsim$4) years, but because they move through the maser shell at comparable velocities, their superimposed emission gives the impression of a single long-living maser cloud. A notable exception was found in RT\,Vir, which showed an individual emission feature (i.e. maser cloud) with a constant (within <0.06 \kms yr$^{-1}$) velocity over 7.5 yr. The maser cloud is assumed to travel in the outer part of the water-maser shell, where the stellar wind apparently has already reached its terminal velocity. For RX\,Boo and U\,Her, we were able to combine our single-dish data with multi-epoch interferometric observations. These maps demonstrated that their maser shells are not uniformly filled, but that there are preferred locations where the conditions are favourable for maser excitation. These regions may persist on timescales of about the wind-crossing time through the \water -maser shell ($\sim 10-20$~years) 

In this paper, we present the results for about two to three decades of monitoring of four more Mira variables that are listed in Table \ref{centralcoords}. Two of these, o\,Cet and R\,Cas, are also part of the MLP programme, and we include data up to and including December 2023.

\begin{table*}
\caption{Basic information on the observed long-period Mira variables. 
}
\label{centralcoords}

\resizebox{\textwidth}{!}{
\begin{tabular}{rlllrrrccccl}
\hline\noalign{\smallskip}
\multicolumn{1}{c}{Name} &  \multicolumn{2}{c}{$\alpha$\,\,\,\, (J2000)\,\,\,\, 
 $\delta$} & 
 \multicolumn{1}{c}{$D^{\rm a}$} &
  \multicolumn{1}{c}{$V_{\ast}$} & 
 \multicolumn{1}{c}{$V_{\rm exp}$} & 
 \multicolumn{1}{c}{$V_{\rm b}$, $V_{\rm r}$} & 
 \multicolumn{1}{c}{$P_{\rm opt}$} &
\multicolumn{1}{c}{$P_{\rm rad}$} &
\multicolumn{1}{c}{$\phi_{\rm lag}$} &
  \multicolumn{1}{c}{Observing}
 & \multicolumn{1}{l}{Notes$^{\rm b}$} \\
\multicolumn{1}{c}{} &
\multicolumn{1}{l}{\, h\,\,  m\, \,   s}
 & \multicolumn{1}{l}{\, \,  $\circ$\,\,\, $\prime$\, \,  $\prime\prime$}
 & \multicolumn{1}{c}{pc} 
 & \multicolumn{1}{c}{\kms}
 & \multicolumn{1}{c}{\kms}
 & \multicolumn{1}{c}{\kms}
 & \multicolumn{1}{c}{days}
 & \multicolumn{1}{c}{days}
 & \multicolumn{1}{c}{} 
 & \multicolumn{1}{c}{Period}
 & \multicolumn{1}{c}{} \\
\hline\noalign{\smallskip}
o~Cet& 02:19:20.8 & $-$02:58:37&92$^{+11}_{-10}$& 46.5&8.0 & 45.6, 48.6& 332 & $336\pm3$ &0.33$^{\dagger}$ & 1990-2023 & 2,3,9,13\\[0.1cm]
R~Leo & 09:47:33.5 & +11:25:44&71$^{+17}_{-11}$&0.0&9.0 &$-$2.2, 1.7 &310 &-- &-- &1988-2011 & 1--5,9 \\[0.1cm]
$\chi$~Cyg & 19:50:34.0 & +32:54:51&160$^{+7}_{-7}$&10.0&8.5 & --& 407 & --&-- &1995-2009 & 2,3,7,9\\[0.1cm]
R~Cas& 23:58:24.7 & +51:23:20&174$^{+6}_{-6}$&26.0&13.5 &22.2, 32.1 & 431 &$430\pm4$ &0.32$^{\S}$ &1987-2023 & 2--7,9 \\[0.1cm] 
\noalign{\smallskip}
\hline
\end{tabular}}
\\[0.1cm]
Notes: We list the distance $D$, the stellar radial velocity $V_{\ast}$, the final expansion velocity in the CSE $V_{\rm exp}$, and the extreme blue $V_{\rm b}$ and red $V_{\rm r}$ velocities of the \water -maser emission. $P_{\rm opt}$ and $P_{\rm rad}$ are the optical and radio period, respectively, and $\phi_{\rm lag}$ is the difference in phase between the two. The entries in the table are discussed in the sections pertaining to the individual stars.\\
References. $^{\rm (a)}$ For distances: o Cet, R Leo: \cite{vanleeuwen08}; 
$\chi$ Cyg, R Cas: \cite{gaiacol20} (EDR3). $^{\rm (b)}$ For stellar systemic and expansion velocities. 1.~\cite{danilovich15}; 2.~\cite{debeck10}; 3.~\cite{devicente16}; 4.~\cite{gon-alfonso98}; 5.~\cite{gon-delgado03}; 6.~\cite{loup93}; 7.~\cite{neri98}; 8.~\cite{sanchez00}; 9.~\cite{teyssier06}; 10.~\cite{chapman94}; 11.~\cite{young95}; 12.~\cite{kerschbaum99}; 13.~\cite{kemper03}. \\
Phase lags:  $^{\dagger}$: calculated for 1995 -- 2011; $^{\S}$: calculated for 1987 -- 1999. \\
\end{table*}

\section{\label{observations} Observations}
Under the Medicina/Effelsberg programme, single-dish observations of the \water -maser line at 22235.08~MHz were made with the Medicina\footnote{The Medicina 32 m "Grueff" VLBI antenna is operated by the INAF-Istituto di Radioastronomia, Bologna}  32 m and the Effelsberg\footnote{The Effelsberg 100 m antenna is operated by the Max-Planck-Institut für Radioastronomie (MPIfR), Bonn} 100 m telescopes at typical intervals of a few months. 
Observations with the Medicina "Grueff" telescope began in 1987 and continued until 2011. For some stars additional spectra were taken in 2015. 
The Effelsberg telescope participated in the monitoring programme between 1990 and 1999. For the details of the observations made within Medicina/Effelsberg monitoring programme  1987 -- 2015, we refer to our previous publications (Paper I -- III).

After several significant improvements were made to the antenna, backends and operating system, monitoring at Medicina was started again in 2018, on a smaller sample and with more frequent observations.
The observations for the MLP reported here were made between May 2018 and December 2023, with an interruption from December 2019 to September 2020 when the telescope was not in use due to maintenance and the COVID pandemic. Spectra were taken approximately once a month.

As a backend we used XArcos \citep{melis15} in
one of its standard configurations (XK00), allowing us to observe left- and right-hand circular polarisations simultaneously, and with progressively higher spectral resolution in the so-called ``zoom mode''.
Spectra are taken in 4 bands that are centred at the rest frequency of 22235.0798~MHz, corrected for the expected LSR velocity of the sources.
The bands have widths of 62.5~MHz, 7.8~MHz, 2~MHz and 0.5~MHz; all have 2048 channels, which lead to resolutions of 30.5~kHz/0.44~\kms, 3.8~kHz/$5.14 \times 10^{-2}$~\kms, 0.95~kHz/$1.29 \times 10^{-2}$~\kms\ and 0.24~kHz/$3.21 \times 10^{-3}$~\kms. In our analyses we used the data from the 7.8~MHz band, smoothed to a resolution of 0.10~\kms.
Observations were done in postion-switching mode, with the OFF-position 0.5\degr\ to the east of the source. The basic observing cycle was composed of four ON and four OFF scans, followed by one CAL scan, each of 30~seconds duration; the cycle was usually repeated three times.
Data were visually inspected in order to flag poor quality spectra. Left- and right-hand polarisation spectra were processed separately and then averaged, to obtain a total of 12 minutes ON-source time. 

Checks of the pointing model were made regularly (typically every $2 - 3$ hours) using quasars. This verified that the pointing accuracy was generally better than $\sim 15$\arcsec. The FWHM of the Medicina K-band receiver is $\approx$ 1\farcm7 at 22 GHz.

Signals were corrected for atmospheric opacity, which was determined from fitting a model atmosphere to skydip observations. Skydips were performed a few times per observing session. A gain correction for elevation was applied, and antenna sensitivity (K/Jy $\sim$ 0.10) was determined by means of cross-scan observations of known flux calibrators (3C123, 3C286, and/or NGC7027), and using flux density values from \cite{perley13}.

\section{Presentation of the data \label{presdata}}
As in the previous publications in this series, for each star we show a selection of the spectra taken over the years, in the sub-sections where they are presented. See Fig. \ref{fig:rcas_sel} as an example. All maser spectra for the stars (except $\chi$Cyg) are presented in the Appendix, available on \href{https://zenodo.org/records/15534987}{Zenodo}. 

To further present and analyse the behaviour of the water-maser emission in time, we again use a number of diagnostic plots, diagrams and light curves, such as the FVt-plot, upper- and lower-envelope spectra, detection-rate histograms and the light curves of the maser spectral components. For a more detailed description of these tools we refer to Papers II and III.

In the following sections we present and discuss the data on the stars in our sample. 
$\chi$~Cyg is only briefly discussed, because a water-maser has never been detected towards this star in our observations.  

Radio light curves are created in general using the integrated flux density determined over a fixed velocity interval. This choice is superior to the use of maximum flux densities of individual emission components because these are not necessarily representative for the general strength as the maser profile varies with time. An exception are simple profiles consisting of a singular feature as that in R~Leo (see below, and Table~\ref{centralcoords}).  
The velocity intervals chosen represent the range of velocities with maser emission shown in the upper-envelope spectrum and detection statistics plots. 

\section{R~Cas}
R\,Cas is a long-period variable AGB star at a distance of $174 \pm 6$ pc (Table \ref{centralcoords}), based on the parallax measured by Gaia. (It was also measured by VLBI, with much larger uncertainties, as 176$^{+92}_{-45}$ pc; see  \citealt{vlemmings03}). We adopt as radial velocity of the star $V_{\ast} = 26.0\pm0.5$ \kms\ and a final expansion velocity $V_{\rm exp} = 13.5 \pm 1.0$ \kms, as determined from circumstellar CO (Table \ref{centralcoords}).

The \water -maser of R\,Cas was first detected in 1971 by \cite{turner71} at 25.9 \kms, which is coincident with the adopted stellar radial velocity. \cite{olnon80} re-detected the maser in 1976 at the same velocity, and regular observations of the maser then started within the Pushchino monitoring programme in 1981 \citep[hereafter P04]{pashchenko04}. They found the maser profile peaking in general in two line-of-sight (los)-velocity intervals $V_{\rm los} = 23-25$ and $25-29$ \kms, and in $1988-1989$ emission was found also at $29-31$ \kms. Between 1981 and 1989 the peak flux densities reached from time to time several hundred Jy, with these bright phases lasting several months. Until 1998 the average brightness level decreased and the maser was not detected thereafter until the end of their monitoring programme in 2003.

\begin{figure*}
\resizebox{17.5cm}{!}{
\includegraphics{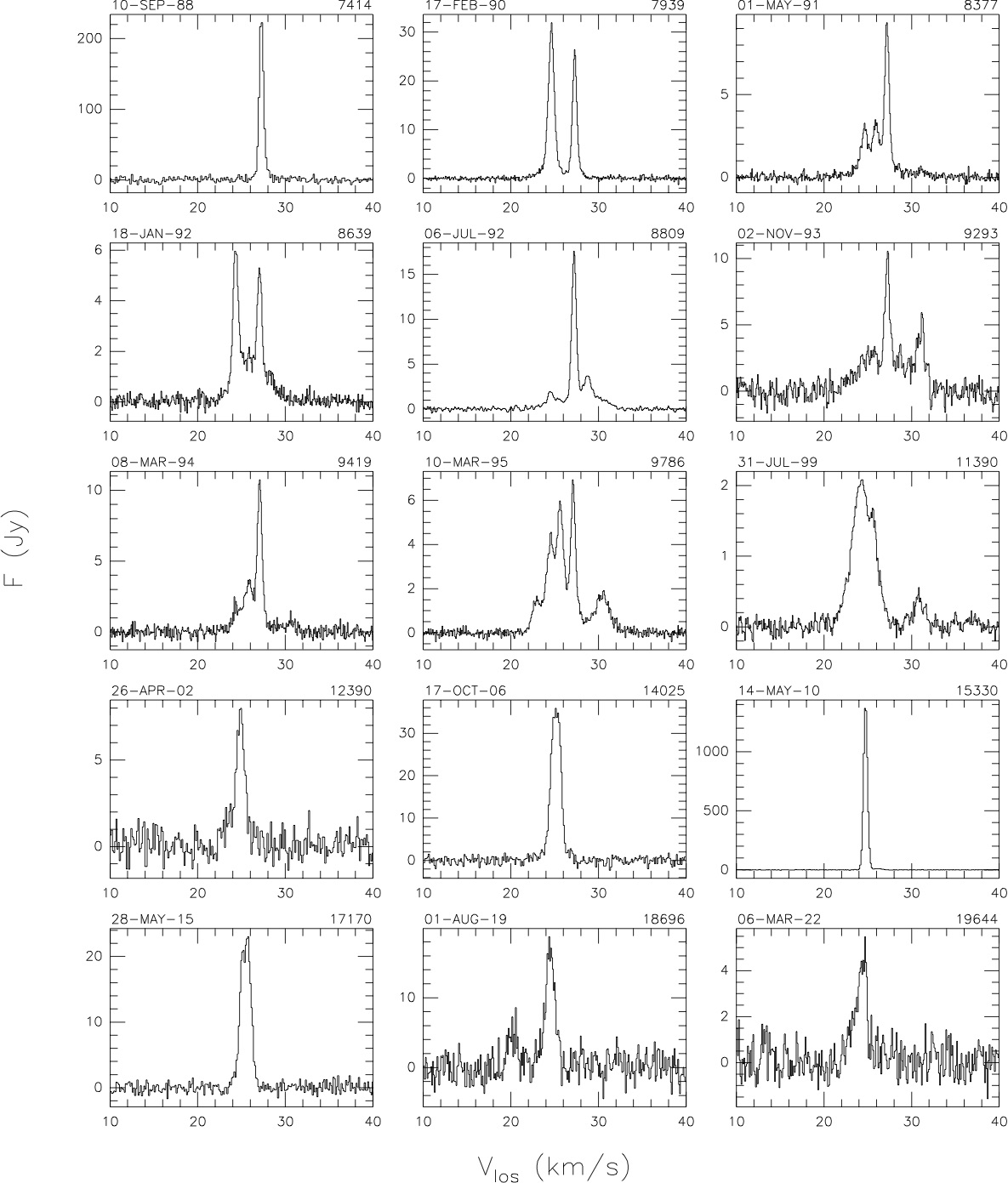}}
\caption{Selected H$_2$O maser spectra of R~Cas. The calendar date of the observation is indicated in the top left corner above each panel, and the TJD (JD-2440000.5) is shown in the top right corner.
}
\label{fig:rcas_sel}
\end{figure*}

Within the Medicina/Effelsberg monitoring programme, observations of R\,Cas were made during 1987 -- 2011 and again in 2015 (a total of 102 spectra), but were only sparsely sampled between the end of 1996 (TJD $\sim$ 10\,400) and the end of 2000 (TJD $\sim$ 11\,800). Our first Medicina spectrum from September 1987 (Fig.\,A.1; Appendix, on \href{https://zenodo.org/records/15534987}{Zenodo}), was already reported by \cite{comoretto90}.
During the monitoring period the maser showed strong variations with peak flux densities between the sensitivity limit (rms was about 0.5 -- 1.5 Jy for Medicina) and more than a thousand Jansky during a burst in early 2010, lasting at least 4 months. In the MLP, R\,Cas is observed more frequently, obtaining a total of 73 spectra in 2018 -- 2023 (with the exclusion of December 2019 -- September 2020; see Section~\ref{observations}). Apart from a brief interval of enhanced activity (TJD $\sim 18500 - 19000$), the maser in R\,Cas has been in a quiescent period since the MLP was started (see Fig.~\ref{fig:rcas-lcurve-three}). To reduce the rms and bring out possible low-level emission in this period, we averaged spectra taken within 4 days, resulting in 58 spectra.
In Fig.~\ref{fig:rcas_sel} representative maser spectra are shown, while the complete set of 160 spectra is shown in Fig.\,A.1 (Appendix, on \href{https://zenodo.org/records/15534987}{Zenodo}).

\begin{figure}
\includegraphics
[width=\columnwidth]
{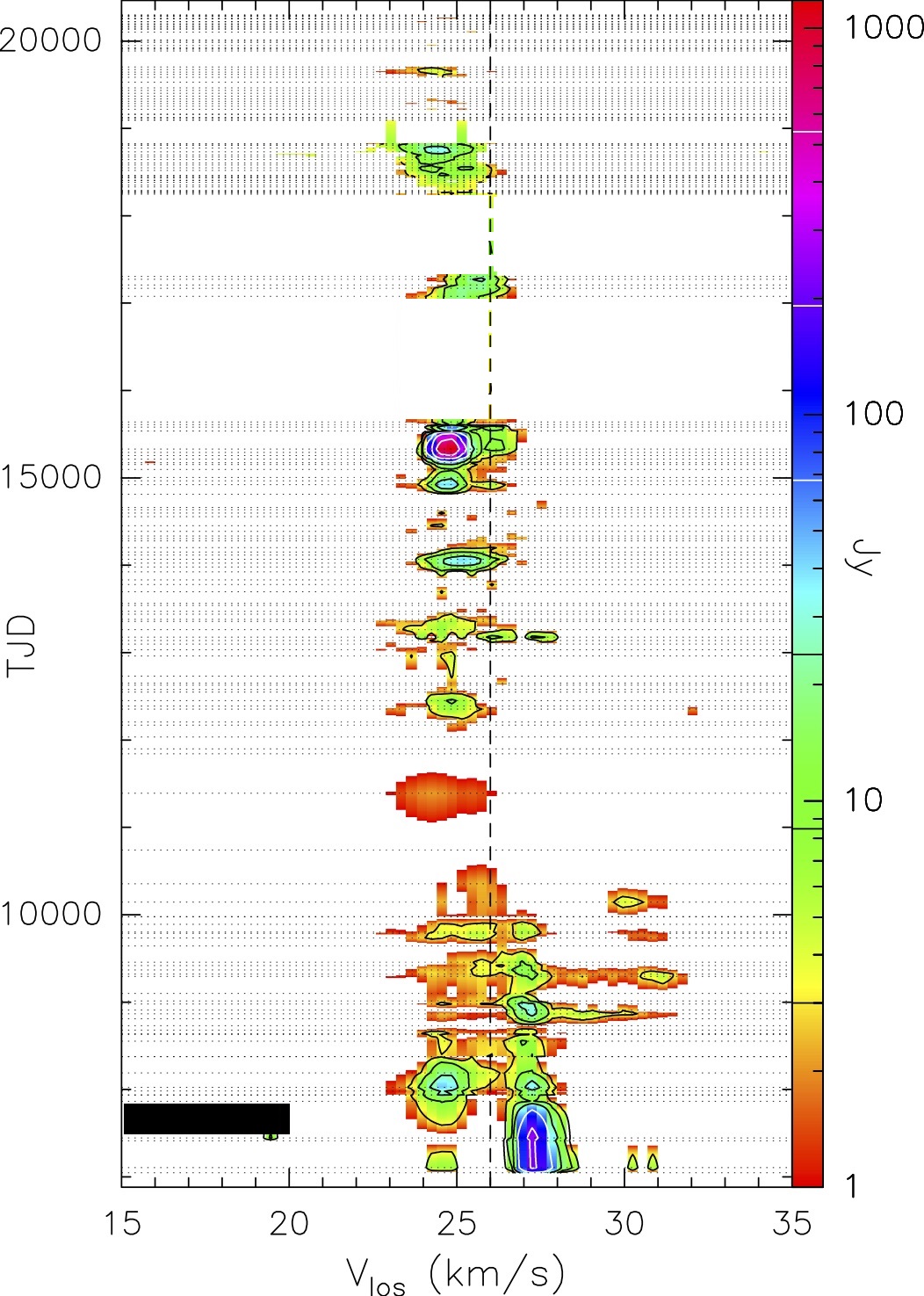}
\caption{Flux density vs. line-of-sight velocity $V_{\rm los}$ as a function of time (FVt-plot) for water-masers in R~Cas. Each horizontal dotted line indicates an observation (spectra taken within 4 days from each other were averaged). The data were resampled to a resolution of 0.3~\kms\, and only emission at levels $\geq 3\sigma$ is shown. The black area in the plot indicates the unobserved parts of the spectrum on these particular days. The first spectrum in this plot was taken on 10 September 1987; JD = 2447048.5, TJD = JD$-$2440000.5 = 7048. The last spectrum in the panel is from 21 December 2023. 
}
\label{fig:rcas-fvt}
\end{figure}

The \water -maser of R\,Cas was mapped by \cite{lane87}  and \cite{colomer00} with the VLA. Lane et al. observed in October 1983 a single maser spot and derived an upper limit for the spot emission region of 100 mas (r$\sim$9 au at 174 pc), while Colomer et al. in 1990 identified 12 maser emission spots distributed in a shell with a radius of 130~mas (r$\sim22.5$ au). The dominant spectral components at this epoch had velocities of 24.7 and 27.2 \kms, coincident with the strong features we observed in May 1990 (see Fig. \ref{fig:rcas_sel}). 

Single-dish observations were made during our monitoring programme of R~Cas by \cite{takaba94, takaba01} and \cite{kim10}. \cite{takaba94} detected 5.5 Jy maser peak emission at 26.1 \kms, between 26 April and 10 May 1991, while we observed another feature to be the brightest with $\sim8$\,Jy at 27.1 \kms\ on 30 April / 01 May 1991. \cite{takaba01} reported for October -- November 1991 a maser peak with 2.7 Jy at 25.8 \kms, while our spectrum from 25 October 1991 has the strongest maser feature at 25.4 \kms\ with a flux density of 10 Jy. \cite{kim10} observed in June 2009 and found a 25~Jy signal at 24.9~\kms; we detected 23~Jy at 24.8~\kms\ on 23 June 2009. There is agreement between the literature and our observations concerning the maser peak velocities. Leaving calibration uncertainties out of consideration, the variations of the peak flux densities indicate brightness fluctuations on the timescale of weeks and, as observed in April/May 1991, apparently also the strongest peak within the profile can change on this timescale. 
 
\subsection{Variations in the \water -maser profile \label{profilevars}}

An overview on the general behaviour of the maser variations is shown in the FVt-plot (Fig. \ref{fig:rcas-fvt}). Until the end of 1995 (TJD $\approx 10\,000$) there are two dominant components, in the ranges $V_{\rm los} = 23 - 25$ and $25 - 29$ \kms\ (see also P04); after that only the former is seen. Also until 1996 emission at 30 -- 31~\kms\ was  detected. There is very weak ($\lsim 0.4$ Jy) emission at velocity 30 -- 32~\kms\ in the Effelsberg spectrum of 31 July 1999 (TJD 11390). More in general, until about 2005 apart from the main components there is occasional low-level emission within the velocity range $V_{\rm b} \leq V_{\rm los} \leq V_{\rm r}$ as given in Table~\ref{centralcoords}. 
In Figure~\ref{fig:rcas-fvt} brightenings of the maser are seen at times, which  occasionally achieve exceptionally strong flux density levels, which we call 'bursts'. The first bursts, with peak flux densities 200 -- 300\,Jy (see Fig.\,A.1 on \href{https://zenodo.org/records/15534987}{Zenodo}), occurred in November 1987, and September/October 1988 (TJD\,$\sim 7000$ and $\sim 7500$); the observations reported in P04, conducted more frequently and covering a longer time interval, show that in a time span of almost  3 years, roughly between September 1986 and June 1989, there were three burst-events of the maser (see also \citealt{pashchenko90}). We registered another burst episode in early 2010 (TJD\,$\sim 15300$) reaching $\sim$1400\,Jy and lasting at least four months. A third time interval  of relatively high brightness occurred in early 2019 and lasted until at least the end of that year. Representative spectra of the bursts are included in Fig. \ref{fig:rcas_sel}. 

\begin{figure}
\resizebox{9cm}{!}
{\includegraphics
{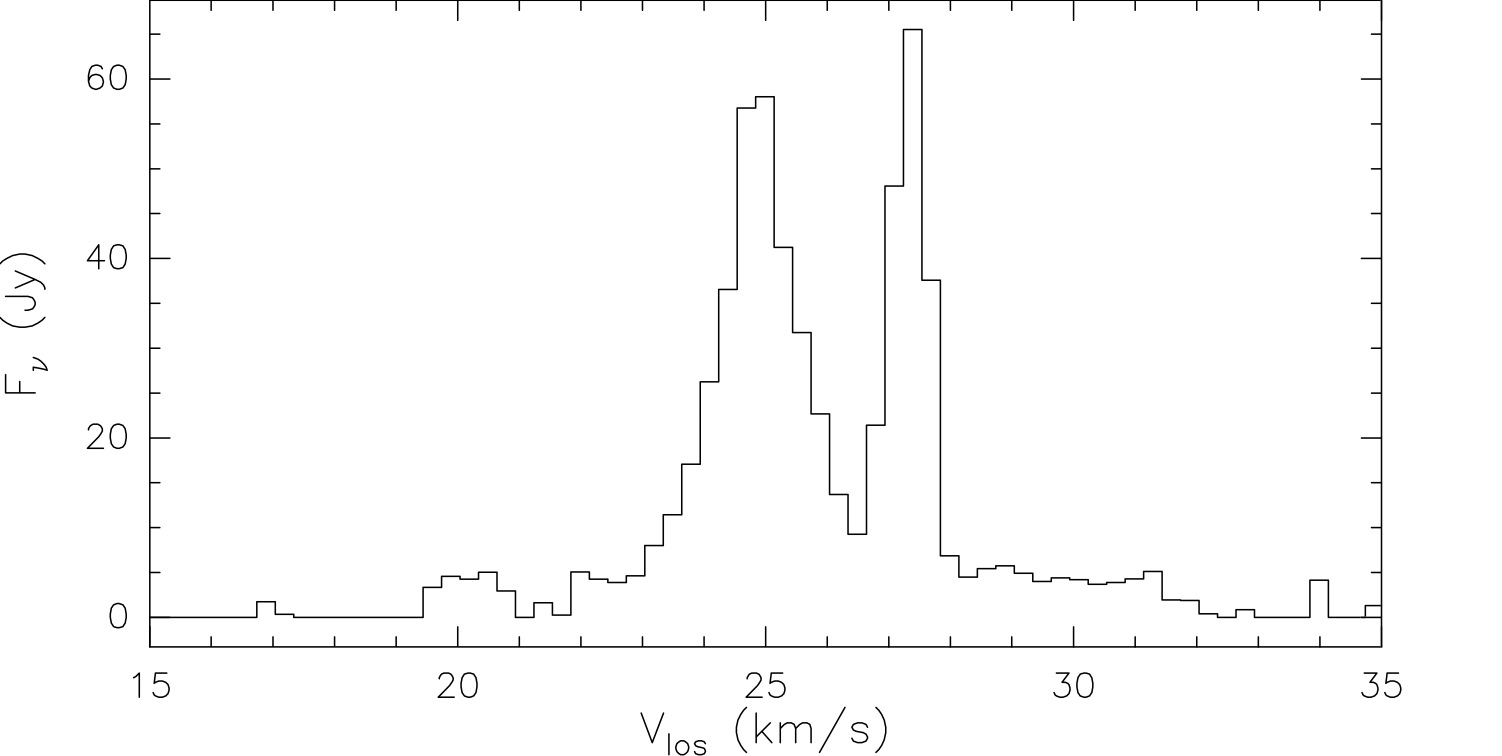}}
\caption{
  Upper-envelope spectrum for water-masers in R~Cas $1987 - 2023$ without the spectra during the major bursts in $1987 - 1988$ and 2010. 
}
\label{fig:rcas-upenv-noflares}
\end{figure}

\begin{figure}
\resizebox{9cm}{!}
{\includegraphics
{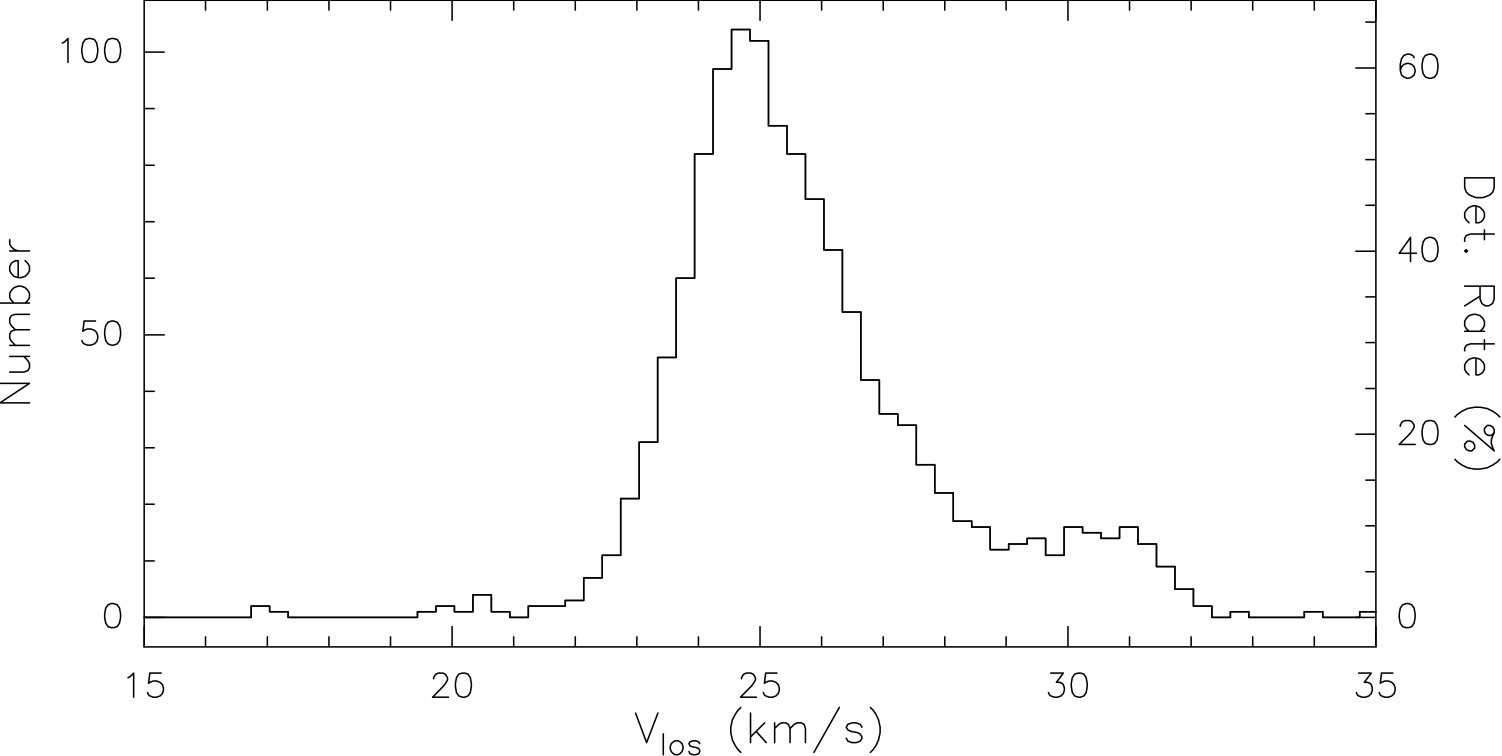}
}
\caption{detection-rate histogram for water-masers in R~Cas from $1987-2023$. 
}
\label{fig:rcas-histo}
\end{figure}

{Figure~\ref{fig:rcas-upenv-noflares} shows the upper-envelope spectrum, excluding 7 spectra involved in the two major bursts in 1987-88 and 2010. As there were times when the maser emission was not detected, the lower-envelope is zero, and is not shown. The upper-envelope spectrum confirms that during our monitoring campaign the maser emission is dominated by two components at 24.5 and 27.5~\kms, respectively. Both components, at separate times, have demonstrated instances of increased maser activity (including the burst spectra affects only the two main components, not the lower-level emission). During the first $\sim$3000 days of our monitoring programme the two dominant components were immersed in a broad, multi-component lower-level emission which we called "the plateau" in earlier publications (e.g., \citealt{brand02}). As of $\sim 1998-1999$ the plateau was no longer detected at our levels of sensitivity, and the 24.5~\kms\ component was the only one left, up to the end of 2023 - see Fig.~\ref{fig:rcas-fvt}. The detection-rate histogram in Fig.~\ref{fig:rcas-histo} shows both the persistence of the 24.5~\kms\ component over the full observing period and the emission over a much wider velocity range seen in the years before 2000. 
To determine the velocity boundaries of the maser emission for $1987 - 2023$ we used  this histogram. Emission at a certain velocity is considered spurious if there are fewer than 5 detections. The boundaries obtained are $V_{\rm b} = 22.2$ and $V_{\rm r} = 32.1$ \kms\  (Table \ref{centralcoords}).

\begin{figure}
\resizebox{9cm}{!}
{\includegraphics
{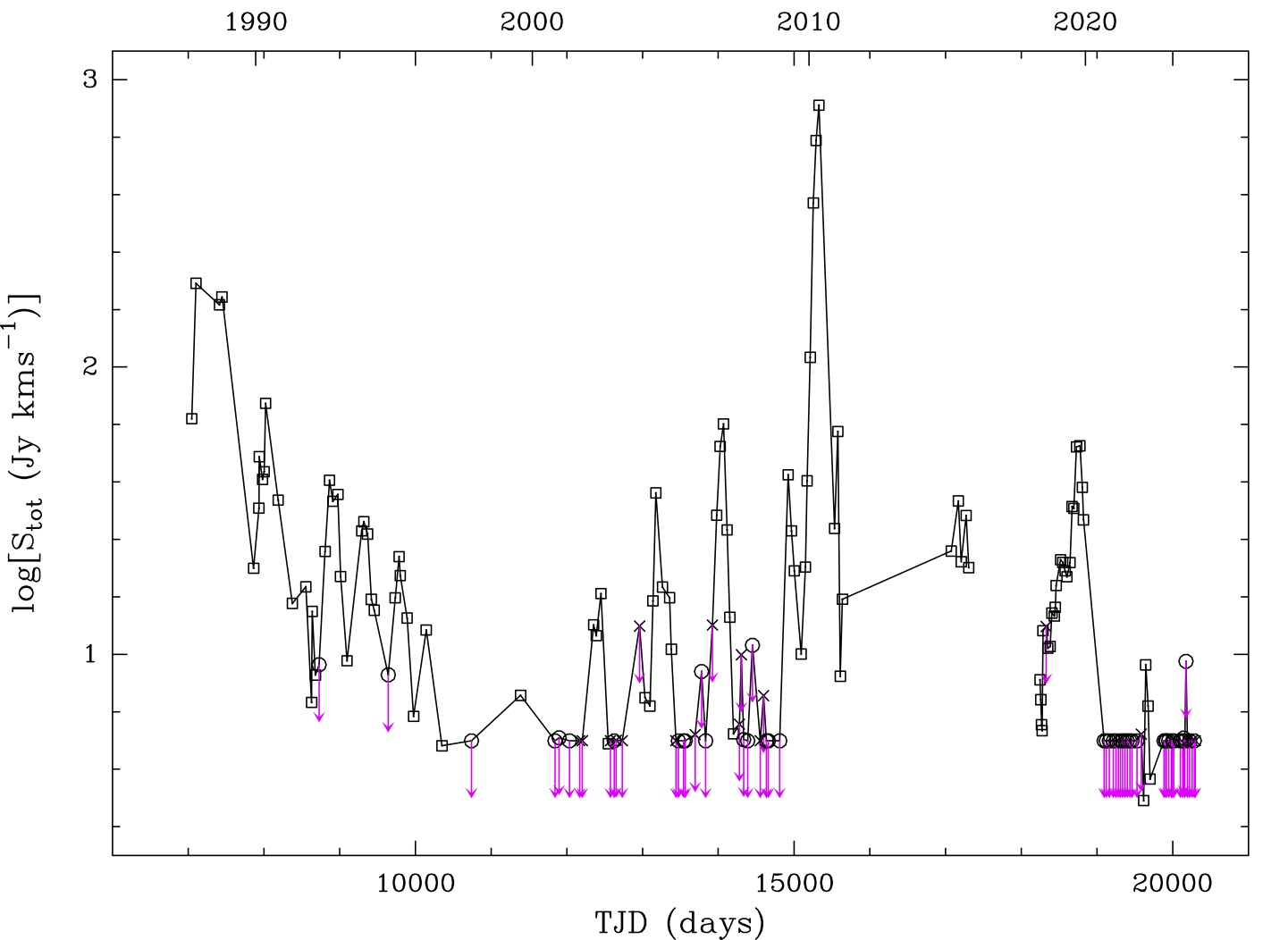}
}
\caption{Logarithm of the integrated emission ($20 \le V_{\rm los} \le 35$~\kms) vs. time between 1987 and 2023 for R\,Cas. The detections are marked with a square, and upper limits are indicated with downward arrows for spectra with marginal (indicated by an x) and non-detections (indicated by an open circle).  
}
\label{fig:rcas-logstot}
\end{figure}

\begin{figure}
\includegraphics[angle=0,width=\columnwidth]
{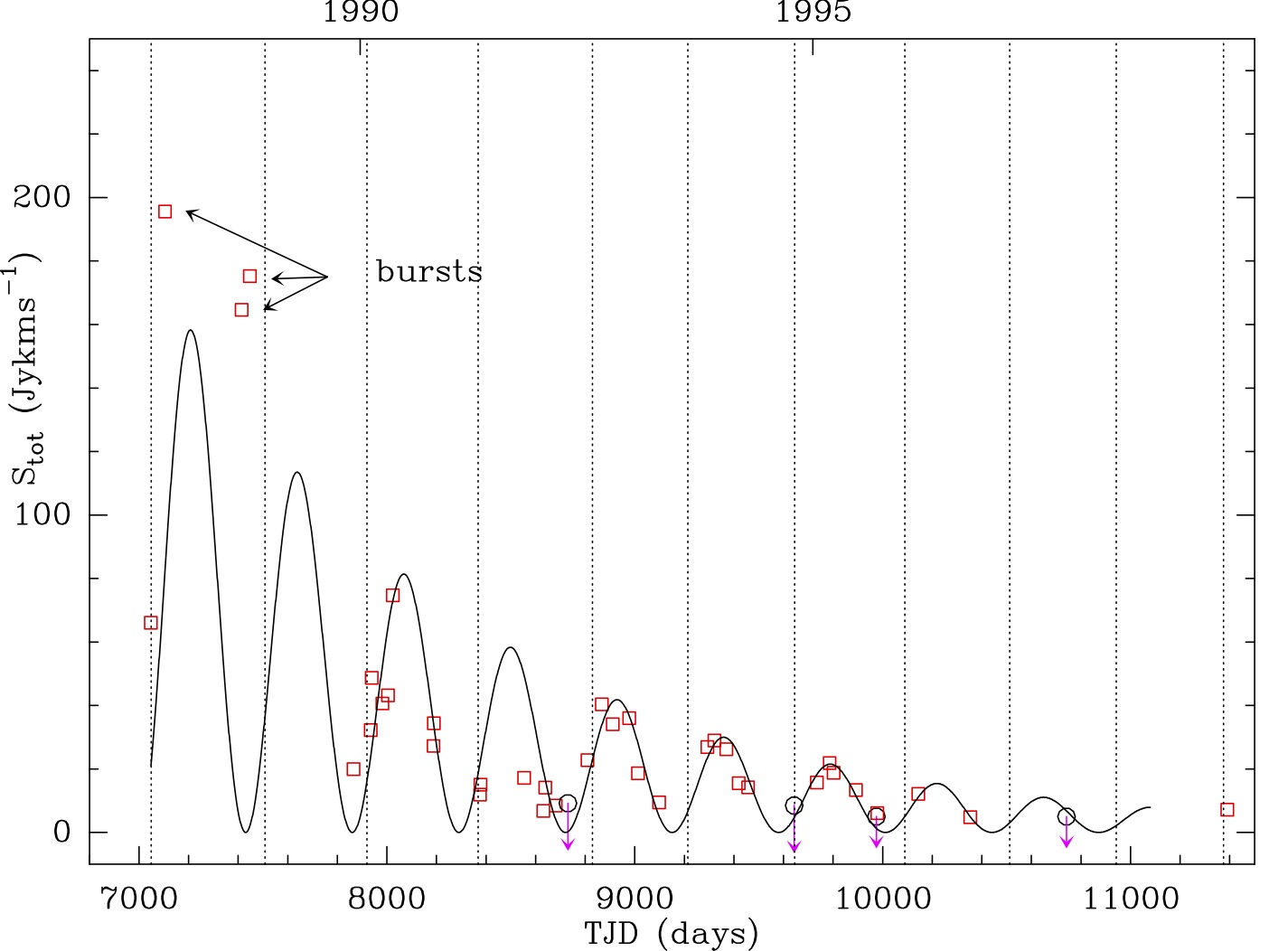}
\caption{ R~Cas \water -maser light curve 1987 -- 1999 displayed in integrated fluxes vs. TJD. The red squares denote detections; the black circles denote non detections; downward arrows indicate upper limits in those cases. The drawn curve shows a damped oscillator fitted to the data and has a period of $430 \pm 4$ days, very similar to the stellar period of visual variation, the location of the maxima of which are indicated by the dotted vertical lines (visual magnitudes; AAVSO-data).
}
\label{fig:rcas-lcurve-two}
\end{figure}

\begin{figure}
\includegraphics[angle=0,width=\columnwidth]
{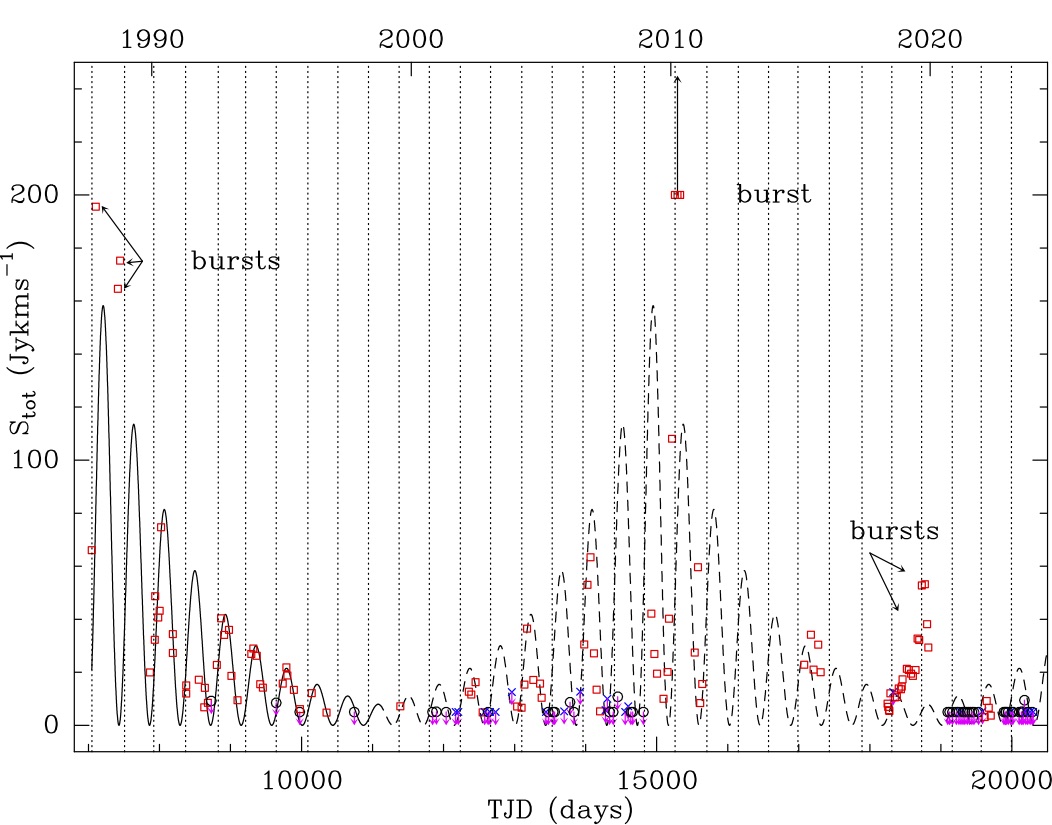}
\caption{ R\,Cas maser light curve in 1987--2023. 
The solid curve shows the damped oscillator fit displayed in Fig.~\ref{fig:rcas-lcurve-two}. The dashed curve between TJD $\approx 11400$ and 15000 is its repetition in reverse direction, while it is repeated in the original direction after TJD$\approx$15000 to show the pattern of repeatability in the data. See the text for discussion. The three observations near TJD 15000, indicated with an upward arrow, were made over 74 days and have $S_{\rm tot}$ of 372, 611, and 814~Jy\,\kms, respectively; they are part of a burst that lasted at least 4 months. The vertical dotted lines indicate the visual maxima of the star (AAVSO data). We distinguish between detections (red squares), non-detections (grey circles) and marginal detections (blue crosses). Based on the pattern in the brightness variations we refer to the intervals $\sim 1987-1997$ and $\sim 1999-2019$ as 'epoch~1' and 'epoch~2', respectively (see text Sect.~\ref{shellkinemat}).
}
\label{fig:rcas-lcurve-three}
\end{figure}

\noindent
\subsection{The maser light curve \label{maserlc}}

The variation in the long-term behaviour of the maser emission is shown in Fig.~\ref{fig:rcas-logstot}, where the integrated flux densities $S_{\rm tot}$ (integrated over the velocity range $20 < V_{\rm los} < 35$~\kms, thus including all emission seen in Fig.~\ref{fig:rcas-fvt}) are plotted against TJD. To create this light curve we classified all spectra as showing either clear-, marginal-, or non-detections based on peak flux density.
Because virtually all non-detections have $S_{\rm tot} < 5$~Jy\,\kms, the minimum upper limit in $S_{\rm tot}$ was set at 5~Jy\,\kms\ for marginal and non-detections, as long as $S_{\rm tot} \le 5$ Jy\,\kms; otherwise the upper limit was set at the observed value. 

The light curve is complex and shows variability patterns on different timescales: bursts, periodic variations and long-term bright and weak phases. To study the periodic variations we used for comparison the visual magnitude data (1987 -- 2023) from the AAVSO-database to determine the dates of visual maxima and the average period $P_{\rm opt}$. For the entire observing time interval covered here, 1987-2023, we find an average optical period (i.e. separation between maxima) of $431 \pm 16$ days.  

Figure~\ref{fig:rcas-lcurve-two} shows} the maser light curve for the first 12 years in close-up. Overlaid are the dates of the visual maxima  (dotted lines).  A few things are noticeable from this plot: first, the total flux decreases with time, while still following the visual variations of the star, and thus, the decrease is not monotonous, but modulated. This period of decreasing activity lasted until 1996, in agreement with the observations by P04. Second, the maser has been observed to be very bright between November 1987 and October 1988, as discussed in Sect.~\ref{profilevars}.
Leaving out these three points, indicated as 'bursts' in Fig. \ref{fig:rcas-lcurve-two}, we fitted a sinusoid, or rather a damped oscillator, to the data between 1987 and 1997, which is shown in the figure (a weighted fit, with weights equal to the inverse square of the uncertainty in $S_{\rm tot}$). We found a period of 430$\pm$4~days (adopted as radio period in Table \ref{centralcoords}) and an exponential damping factor exp($\alpha$t) with $\alpha = (-7.73 \pm 0.67) \times 10^{-4}$; in this same time interval the AAVSO-data indicate an average optical period of $430 \pm 24$ days. 
The delay of the maser maxima in 1987 -- 1999 with respect to the optical maxima is about 135~days ($\phi_{\rm lag} = 0.32$), larger than the delay $\phi_{\rm lag} \sim 0.2$ measured in the near-infrared in 1988 -- 2000 by \cite{nadzhip01}. We assumed that both values are compatible within the uncertainties, however. 

Around 1997 a weak phase sets in, during which P04 could not detect the maser either. This phase is not well covered with observations, in fact until October 2000 only two observations were made. Since then (TJD $>$ 11800) the regular observation in Medicina resumed and the maser was not detected until October 2001. A single detection was made during this 'weak phase' in Effelsberg in July 1999 with a peak flux density of 2 Jy (TJD = 11390). It seems likely that the maser was not completely extinguished during the 'weak phase' of almost 5 years, but fluctuated on flux density levels which rarely (or never) exceeded $\sim2$ Jy.

Figure~\ref{fig:rcas-lcurve-three} shows the full set of data obtained during our monitoring campaign, from 1987 to 2023. with two long interruptions lasting 4 and 2.5 years. respectively ($2011 - 2015$, $2016 - 2018$). The weak phase is seen to have lasted until about 2002, after which there was another active phase that continued until observations were interrupted in 2011.
In this phase, the maser consisted of a single feature at $\sim 24.5$ \kms\ 
and times with clear detections alternated with non-detections\footnote{In April 2003 observations by \cite{vlemmings05} with the VLBA failed to detect the maser. As they already suspected this was because of the weakness of the maser in this particular month (see Fig.\,A.1 on \href{https://zenodo.org/records/15534987}{Zenodo}).}. Between April 2007 and December 2008 there were only marginal detections at best. 
In January to May 2010 a strong burst was observed, reaching a peak flux density of $\approx$1400~Jy in May 2010. 
This time the burst occurred in the 24.5~\kms\ component only. The rise of the maser during the burst to its peak flux density happened in parallel with the optical brightening of the star, but occurred shortly after the maximum optical light. The actual radio maximum may have been missed, because the radio light curve lags 
0.32 in phase behind the optical light curve. The decline of the burst was not covered; at the next observation in December 2010 
it had vanished. Unlike the burst observed in for example R~Leo (see Sect. \ref{sec:rleo}) the duration of $<1$ year is shorter ($<0.8 P_{\rm opt}$) than the pulsation period of the star. From February 2011 and again in 2015 a dominant maser feature at 24.9 and 25.1 -- 25.6~\kms, respectively, was observed at flux density levels 10 -- 15~Jy (2011) and 20 -- 30~Jy (2015).
Observations were resumed in May 2018, when the maser once more seemed to be entering a weak phase. 
In Fig.~\ref{fig:rcas-lcurve-three} we note a repeating pattern, where the integrated maser emission goes from a strong phase, slowly descends into a quiescent state, then again increases to show a period of high(er) maser activity, and then back to quiescence once more, all the while showing a variation in parallel with the optical variation of the star. We have again drawn the damped oscillator that resulted from a fit to the data between 1987 and 1997, extending it to 1999. We then plotted that curve in the reversed direction after TJD=11080, and then again in the original direction after TJD$\approx$15100. This particular sinusoid fitted to the first decade of data also follows the next two decades fairly well, suggesting that this is a repeating phenomenon. By the end of 2023 the \water -maser of R\,Cas was still faint and its average brightness is expected to increase in the coming years, and we continue monitoring it in the MLP.

In Papers I and III we have found the existence of regions in the CSE where the conditions are favourable for the excitation of (water) masers over many pulsation cycles of the star. In the particular case of RX\,Boo, featured in Paper I, interferometric observations indicated that such a favourable region existed for at least 11 years; for U\,Her (Paper III) such a region was found to be persistent for at least 6.5 years. The existence of such regions was also deduced by, e.g., \cite{rudnitskii90} and P04 (the so-called quasi-stationary layer). In this context, 
one possible explanation of Fig.~\ref{fig:rcas-lcurve-three} is that we see a sequence of mass loss events, with one or more maser clouds entering the part of the CSE of R\,Cas where conditions are right for H$_2$O maser excitation. At first the masers have low emission levels. As a cloud travels through the shell it crosses a zone particularly favourable for maser emission, at which time $S_{\rm tot}$ increases. When it leaves that zone and moves on towards the outer edge of the water-maser shell, its flux density decreases once again until the maser becomes invisible. Projecting this interpretation on Fig.~\ref{fig:rcas-lcurve-three} would mean the appearance of one or more maser clouds around the year 2000 
(TJD $\sim$ 11800), and the occurrence of an event of enhanced mass loss some time before that. Those clouds pass through the 'favourable' maser zone around the year 2009 and disappear more or less in 2021, i.e. the crossing time is $\sim$20 years. At the start of our monitoring campaign in 1987 we then would have caught clouds, consequence of an earlier mass loss event, just as they were leaving the favourable zone on their way out of the shell. The actual situation may be somewhat more complex, as we shall see. \\

\subsection{Shell kinematics \label{shellkinemat}}

\begin{figure}
\resizebox{9cm}{!}
{\includegraphics
{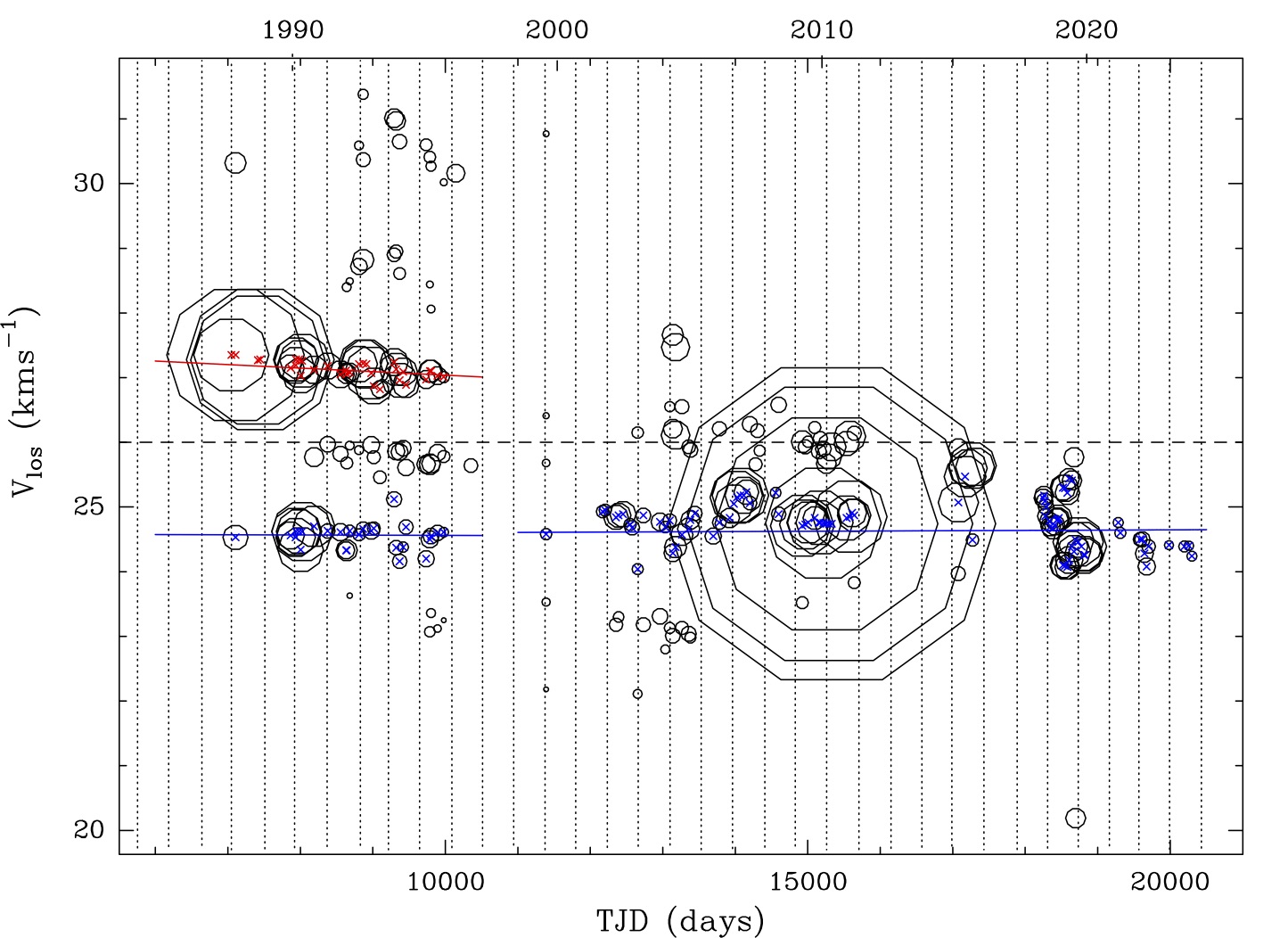}
}
\caption{Water-maser components in R\,Cas, whose peak flux density velocities $V_{\rm los}$ from Gaussian fits to maser emission components, are plotted against time. Sizes of the symbols are proportional to the peak flux density of the components. Solid blue and red lines are weighted least-squares fits to the components at $\sim$24.5~\kms\ and $\sim$27.5~\kms, respectively. Fits are made separately for epochs~1 and 2 (see Fig.~\ref{fig:rcas-lcurve-three}). Components that participated in the fits are indicated by red (27.5 \kms\ component) or blue (24.5 \kms\ component) crosses. The horizontal dashed line marks the stellar radial velocity $V_*$; the vertical dotted lines indicate the star’s visual maxima (AAVSO data). 
}
\label{fig:rcas-fitted-comps}
\end{figure}

\begin{figure}
\resizebox{9cm}{!}
{\includegraphics
{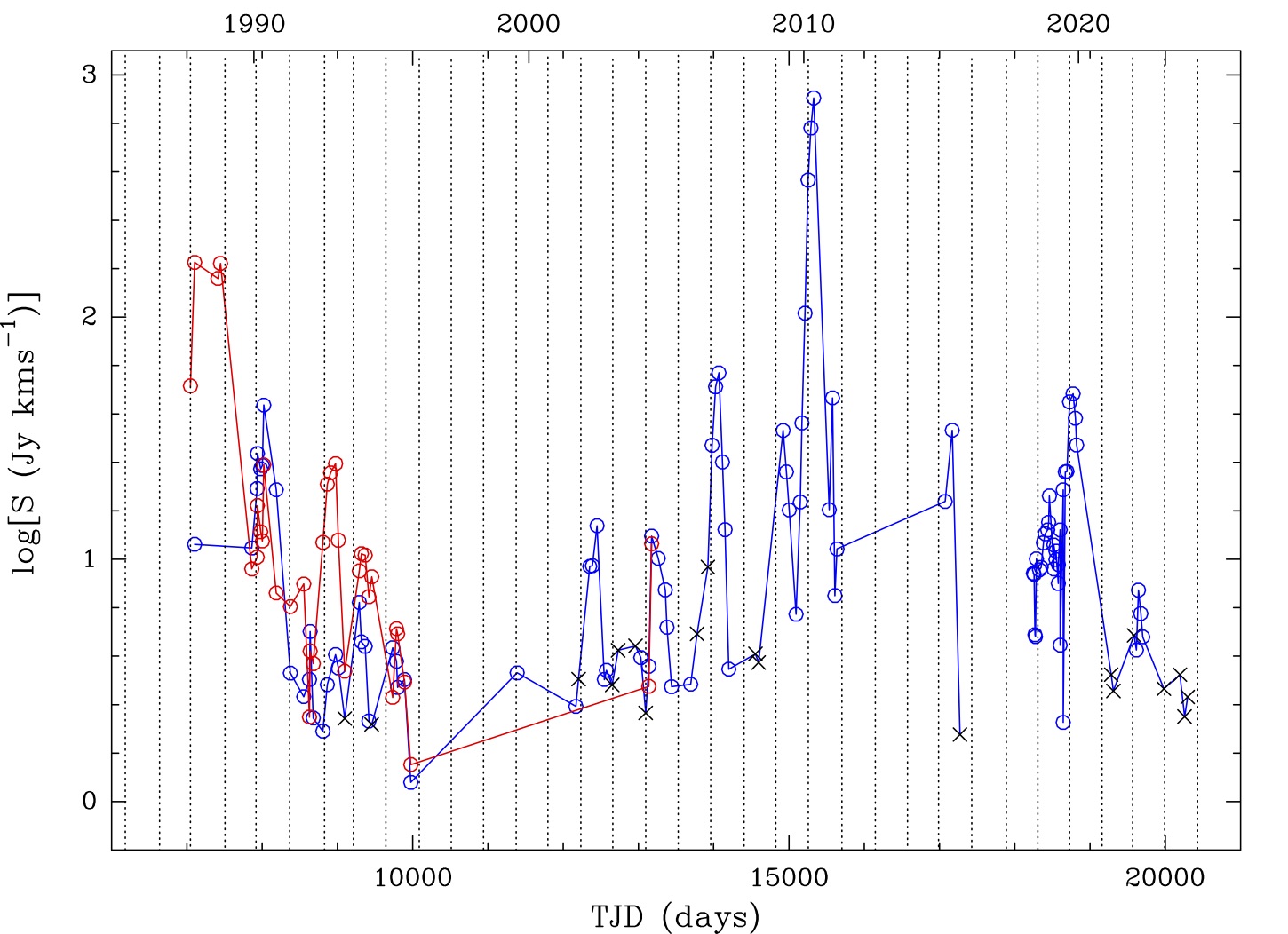}
}
\caption{Logarithm of the velocity-integrated flux of water-masers in R\,Cas, which is derived from the area under Gaussian fits to maser main emission components versus time: $\sim$24.5~\kms\ (blue) and $\sim$27.5~\kms\ (red), respectively. Where the emission line in question has a signal-to-noise ratio lower than 3, the data point is marked with a black cross. When detected, both components demonstrate the same trend in their flux density with time. The vertical dotted lines indicate the star’s visual maxima (AAVSO data).
}
\label{fig:rcas-logs-main-comps}
\end{figure}

To study the maser emission in more detail, we have fitted Gaussians to all emission in spectra with maser detections. The results are shown in Fig.~\ref{fig:rcas-fitted-comps}, in the form of an alternative (and more sensitive) FVt-diagram. The line of sight velocity $V_{\rm los}$ of the Gaussian peaks are plotted against time; the strength of the emission is represented by the size of the symbols (area proportional to flux density). Three components stand out, in the sense that there is recurrent emission at three velocities: 24.5, 26.0 (=$V_{\ast}$) and 27.5~\kms. The components of interest here are those that exhibited the major bursts.
In Fig.~\ref{fig:rcas-logs-main-comps} we show the integrated flux density under the Gaussians of those components. During the first decade of our monitoring programme the 27.5~\kms\ (red) component dominated the spectral profile and the bursts in the interval $1986 - 1989$ are all at that velocity (see also P04). The 24.5~\kms\ (blue) component has also been detected, and when present follows the same periodic variation, but at a lower level (note that the scale in Fig.~\ref{fig:rcas-logs-main-comps} is logarithmic). After the quiescent period of the maser ($\sim 1997 - 2002$), the red component is no longer present and it is the blue component that dominates.

We have fitted straight lines to the line-of-sight velocities of the red and blue emission components, derived from the Gaussian fits, and drawn them in Fig.~\ref{fig:rcas-fitted-comps}.  
We have distinguished two epochs of long-term brightness variations of the maser emission in the CSE of R\,Cas during our monitoring: $1987 - 1997$ (epoch 1) and $1999 - 2019$ (epoch 2).

The slopes of fitted lines are $(-0.311 \pm 2.488) \times 10^{-5}$ \kms\,day$^{-1}$ and $(0.471 \pm 0.736) \times 10^{-5}$ \kms\,day$^{-1}$ for the 24.5 \kms\ component of epochs 1 and 2, respectively, and $(-5.463 \pm 1.261) \times 10^{-5}$ \kms\,day$^{-1}$ for the 27.5 \kms\ component in epoch 1.
While the blue component has a constant $V_{\rm los}$ within the uncertainties, that of the red one shows a significant gradient, but is seen only during epoch~1.

In Fig.~\ref{fig:rcas-fitted-comps} we see small shifts back and forth in the line-of-sight velocity of the 24.5 \kms\ component on timescales of months. As we argued in paper III, such small shifts can be the result of blending caused by different maser clouds with similar velocities and unrelated brightness fluctuations. In fact, in the June 1990 interferometric map made by \cite{colomer00} there are three bright components (6.4 -- 23.4~Jy) that make up the $\sim$24.5~\kms\ component.  These small shifts may reflect the random formation and disruption/disappearance of contributing maser clouds or a variation of excitation conditions on these timescales. 

In the standard model, maser clouds are moving outward through the CSE in an accelerating stellar wind (\citealt{hoefner18}), so if we are seeing the same cloud at all times then we would expect it to gradually change its line-of-sight velocity; similarly for masers originating in cloudlets at different locations along a radius in the CSE at increasing distances from the star. The drift of the 24.5~\kms\ velocity component is practically zero, however. This would happen if the component moves in the plane of the sky, as we conclude from our analysis below for the R\,Cas components. If not, this could either indicate that the emission appears with time at more or less the same location in the shell, or that the acceleration of the stellar wind is very small, or that the maser shell is already outside the acceleration zone (see the ``11~\kms\ component'' in RT\,Vir in Paper II). To distinguish these cases, repeated interferometric observations on a roughly monthly timescale would be required. This way longer-living components moving tangentially outwards in the shell could be followed and/or the coming and going of short-lived components in an excitation-bounded stationary region of the shell could be traced.

The gradient of the 27.5~\kms\ component is the only one significantly different from zero, but it is negative, implying it brings the maser cloud closer to the stellar systemic velocity with time (by $\lsim 0.2$~\kms\ over 10~years). If the emission at different times does indeed come from the same maser cloud, this would indicate infall rather than outflow, or a non-isotropic CSE with a complex velocity field. From interferometric observations made in June 1990, \cite{colomer00} have modelled the distribution of water-maser spots as a shell centred on the 27.5~\kms\ component and coinciding with the position of the star. Based on this and on the fact that the $V_{\rm los}$ of this component is more positive (redder) than the stellar systemic velocity, P04 interpret the emission at this velocity as amplification of the radio continuum emission from the star by masing water molecules in material falling back from the maser shell towards the star, 
for example from the the inner boundary of the shell (see discussion in \citealt{rudnitskii90}). 
However, because the 27.5~\kms\ component is redshifted with respect to the stellar velocity, the material in which this (red) maser component originates probably is on the other side of the star and therefore does not amplify the stellar continuum.
Because our fit (Fig.~\ref{fig:rcas-fitted-comps}) shows that its $V_{\rm los}$ is gradually becoming more blue (negative gradient), until the maser disappeared at around 1997, this material is likely being pulled back towards the star.

\begin{figure*}[t]
\resizebox{18cm}{!}{
\includegraphics{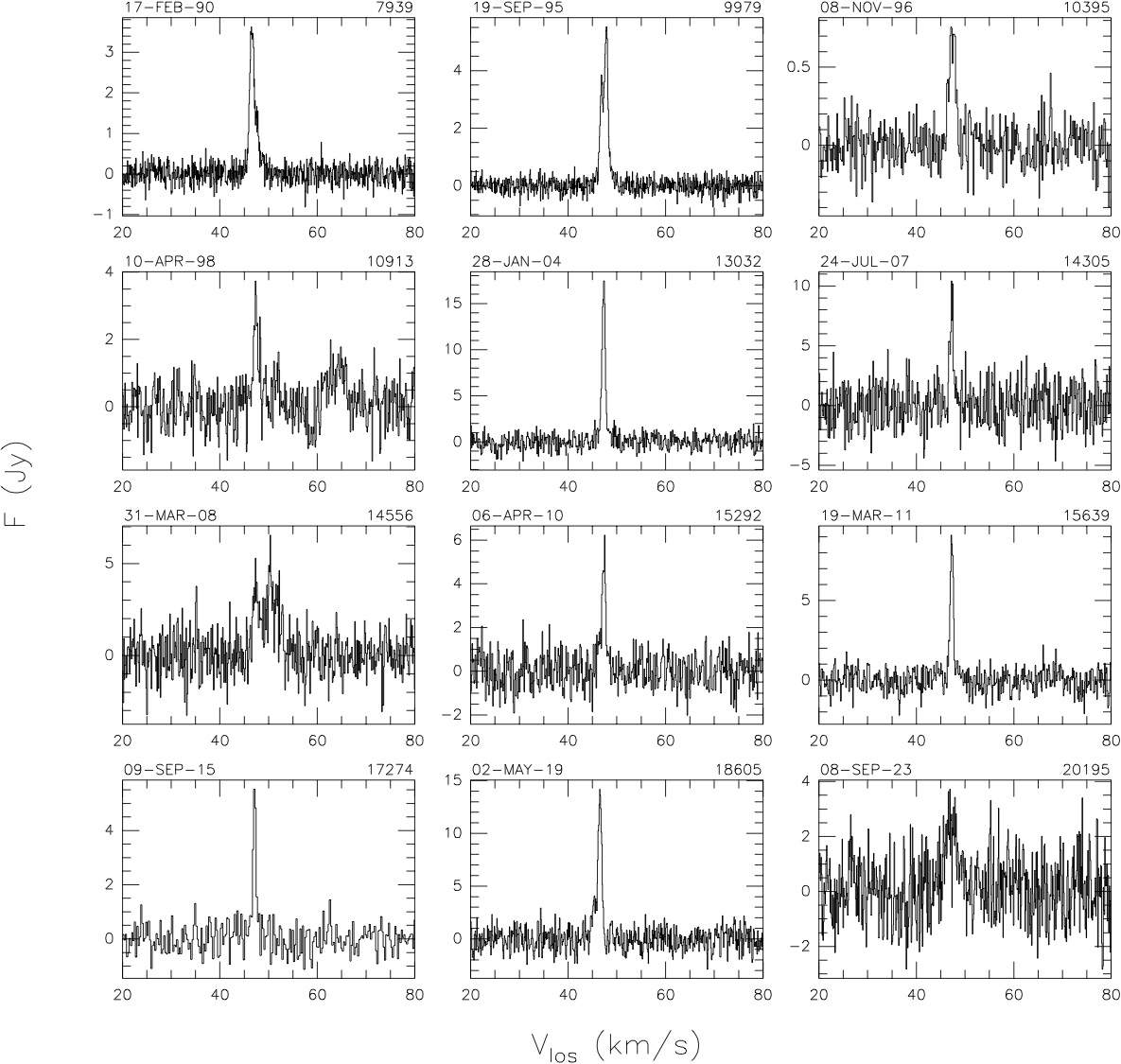}}
\caption{Selected H$_2$O maser spectra of o~Cet. The calendar date of the observation is indicated in the top left corner above each panel, the TJD (JD-2440000.5), in the top right corner. 
}
\label{fig:ocet_sel}
\end{figure*}

We can estimate a range for the size of the shell in the CSE where conditions are such that they allow water-maser excitation. The observed radial velocity of the cloud with maser emission, $V_{\rm los}$, is related to its actual velocity, $V_{\rm out}$, via the angle $\theta$ between the path of the cloud and the line of sight through $V_{\rm out} = \mid V_{\rm los} - V_{\ast} \mid/{\rm cos}\theta$. The 24.5~\kms\ component has a velocity of 1.5~\kms\ with respect to the systemic velocity $V_{\ast}$. We consider the two limiting situations: If its motion is along the line of sight ($\theta=0$) then $V_{\rm out}$=1.5~\kms; if this maser cloud is not moving along the line of sight and at the maximum outflow velocity $V_{\rm exp}$=13.5~\kms\ (Table~\ref{centralcoords}), then $\theta \approx 84^{\degr}$. In the shell crossing time of $\sim$20 years (Sect. \ref{maserlc}), assuming that the velocity of the cloud $V_{\rm out}$ remains constant and $1.5 \le V_{\rm out} \le 13.5$~\kms, the cloud would travel a distance of $\sim 6 - 57$~au, respectively. 

In the maser shell model of \cite{colomer00} the $\sim$24.5~\kms\ component lies at a projected distance of about 100 mas (17.35~au) from the star, which is assumed to coincide with the 27.5~\kms\ component. We argued above, however, that the 27.5~\kms\ component originates from the CSE on the far side of the star and therefore only coincides with the star in projection, if at all. As both of the main components reached a maximum in the years $1987 - 1990$ (see P04 and Figs.~\ref{fig:rcas-fitted-comps} and \ref{fig:rcas-logs-main-comps}) it is plausible to assume the emitting clouds were near the centre of the favourable zones in their part of the shell, and moved out of this shell in the next 10 years; this explains their common decline in flux density in the first ten years of our observations (Fig.~\ref{fig:rcas-logs-main-comps}). From their mutual projected distance (17.35~au) in 1990 we estimate that each component lies at a projected distance of $\sim 9$~au from the star; their true distances are $\sim$ 9/sin($\theta$). Typical radii of the water-maser shells of Mira variables are about 10 -- 30~au (paper III), which implies that for a true distance $\leq 30$~au, $\theta \geq 17^{\degr}$. 
Earlier we found that because for both components $\mid V_{\rm los} - V_{\ast} \mid = 1.5$~\kms = $V_{\rm out} \times$ cos($\theta$) $\leq V_{\rm exp} = 13.5$~\kms, $\theta \leq 84^{\degr}$. We can narrow down this range in $\theta$ by using the velocity curve derived for Mira variable U\,Her in Paper III. For distances $r = 9$ and $30$~au from the star, $V_{\rm out} \approx 5$ and $\approx 11$~\kms\ respectively. The observed $V_{\rm los}$ then implies that $73^{\degr} \leq \theta \leq 82^{\degr}$. To make this consistent with their projected distance of $\sim$9~au, the true distance of either component from the star in 1990 would have been no more than 9.5~au.

The most likely scenario is that around 1990 the positions of the 24.5~\kms\ and the 27.5~\kms\ maser components were at $\sim 9$~au from R\,Cas, in opposing hemispheres with respect to the plane of the sky. They were moving outward in the CSE at a velocity $V_{\rm out}$ of about $4-6$~\kms\ at an absolute angle of about $\theta \approx 70^{\degr}-80^{\degr}$ (24.5~\kms\ component) and $\theta \approx 100^{\degr}-110^{\degr}$ (27.5~\kms\ component) with respect to the line of sight. At an average velocity of 5~\kms\ the maser clouds travel a distance of $\sim$20~au during the crossing time of 20 years; hence the width of the water-maser shell is about this size. In 1990 the clouds already had travelled half of this distance, implying that the outer edge of the \water -maser shell lies at about 20~au from the star.

\section{o~Cet \label{sec.ocet}}
  
o~Cet (Mira) is the prototype Mira variable AGB star with a period of 332 days. It is located at a distance of mere $92\pm11$ pc. We adopt a stellar radial velocity $V_{\ast} = 46.5$ and a final expansion velocity $V_{\rm exp} = 8.0$ \kms, as determined from circumstellar CO and thermal SiO emission (see Table~\ref{centralcoords}). According to \cite{etoka17} the radial velocity is poorly constrained so that the error is about $\pm1$ \kms. Sensitive ($\sigma=0.8$ mJy\,beam$^{-1}$) ALMA observations of \cite{wong16} found $^{28}$SiO $v=0, J=5-4$ emission over a  velocity range significantly larger than previous CO measurements. With a range of 23 \kms, an expansion velocity of 11 -- 12 \kms\ is implied.  \cite{hoai20} reported that the wide velocity range is limited to the inner part of the CSE, however, and may be better explained by the turbulent motion induced by shock waves within $\sim10$~au than by tracing  stellar wind velocities farther out. A final expansion velocity $V_{\rm exp} = 8.0$ \kms\ was therefore kept for consistency with the $V_{\rm exp}$ determinations for the other stars listed in Table~\ref{centralcoords} and those analysed in Papers I -- III, which are based mainly on CO profiles from transitions with low excitation temperatures. 

The \water -maser of o~Cet was discovered in 1973 by \cite{dickinson76} as a single feature at 46.7 \kms. Subsequent observations until the beginning of our observations on a regular basis in 1995 either recovered this feature peaking at a velocity within 46 and 48 \kms\ (\citealt{olnon80}; \citealt{takaba01}), or reported non-detections  (\citealt{deguchi89}; \citealt{comoretto90}; \citealt{takaba94}). Maps were taken with the VLA in 1983 \citep{lane87}   and 1988 \citep{bowers94}, from which  consistent diameters of the  \water -maser region of $4 - 5$~au were derived. Bowers \& Johnston conclude that the maser region consists of at least two spatial components with velocities 46.7 and 47.3 \kms.

In Fig.~\ref{fig:ocet_sel} representative maser spectra from observations taken between 1990 and 2023 are shown, while the complete set of 128 spectra are displayed in Fig.\,A.2 (Appendix, on \href{https://zenodo.org/records/15534987}{Zenodo}). The observations were well-sampled with a $\sim 2 - 4$ month cadence during the time ranges $1995 - 2011$ (but with only two observations between May 1999 and May 2001) and $2018 - 2023$. In addition, a few observations were made in 1990/91 and in 2015. In all those observations the maser showed basically the same single feature profile, which rarely surpassed 10 Jy in flux density, at $V_{\rm los} = 46 - 49$ \kms\ (hereafter 47\,\kms\ feature). The exception is the appearance of additional emission over the velocity interval $\sim$ 48 -- 53 \kms, detected marginally in December 2007 (TJD = 14455) and at the 4 -- 7$\sigma$ level in March 2008 (TJD = 14556) (Fig.\,A.2 in the Appendix on \href{https://zenodo.org/records/15534987}{Zenodo}). This is within the velocity range observed for SiO-maser emission in o~Cet \citep[see below]{jewell91}, but as this emission was seen clearly only in a single spectrum it could be spurious, because of its weakness. The spectrum taken during the maximum in autumn 1995 (Fig. \ref{fig:ocet_sel}) unambiguously shows that the 47\,\kms\ feature is actually a blend of at least two maser lines, with velocities at that time of 46.9 and 47.8 \kms\

\begin{figure}
\resizebox{9cm}{!}{
\includegraphics
{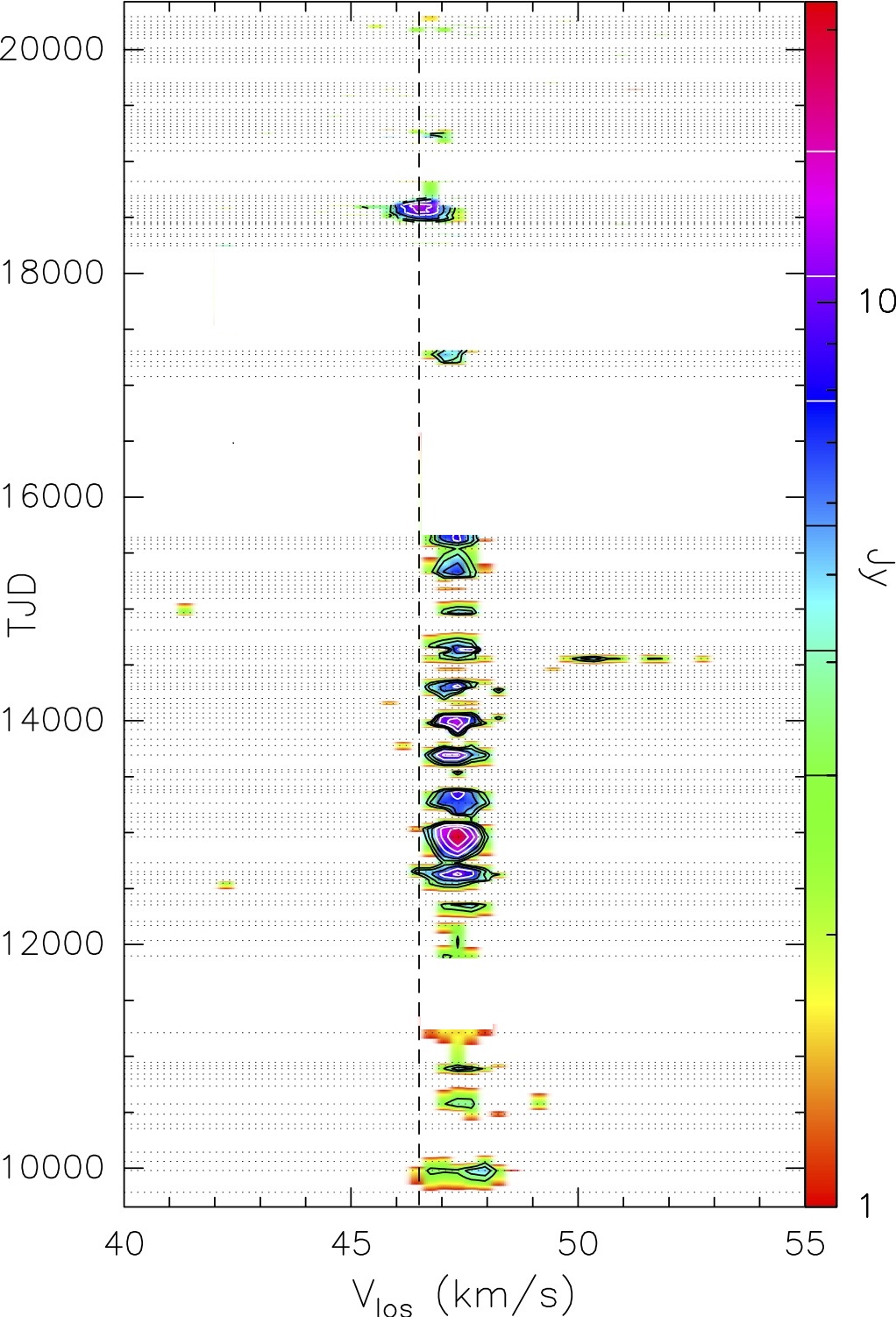}}
\caption{As Fig.~\ref{fig:rcas-fvt}, but for o~Cet. The first spectrum in this plot was taken on 10 March 1995; JD = 2449786.5, TJD = JD$-$2440000.5 = 9786 and the last spectrum on 17 December} 2023.
\label{fig:ocet-fvt}
\end{figure}

An overview on the general behaviour of the maser variations is shown in the FVt plot (Fig. \ref{fig:ocet-fvt}). The maser was regularly detected in 1990 -- 2011 (TJD $\sim$9700 -- $\sim$15700) during the optical bright phases with varying maximum maser flux densities. In the years 2018 -- 2023 the \water -maser of o\,Cet was on average weaker than in the years 1990 -- 2015. Except for the first six months in 2019 and in January/February 2021 the maser was not detected (rms $\sim$ 1 Jy) or marginally present with flux densities of a few Jy (but no observations were made between December 2019 and September 2020, see Sect.~\ref{observations}). In 2019 and 2021 the maser consisted of a single feature at a peak velocity of $46.55\pm0.10$ \kms\ and $46.75\pm0.10$ \kms, respectively. In parallel with our observations, o~Cet was observed in the VERA-Iriki monitoring programme (Shintani et al. 2008) between September 2003 and October 2006 with revisits every six weeks on average.  With a sensitivity of $\approx$1 Jy similar to ours, they detected emission at $V_{\rm los} \sim 47.5$ \kms\ also only during the optically bright phases of the stellar brightness variations. In 2009 \cite{kim10} detected the maser also at the same velocity.

\begin{figure}
\resizebox{9cm}{!}
{\includegraphics
{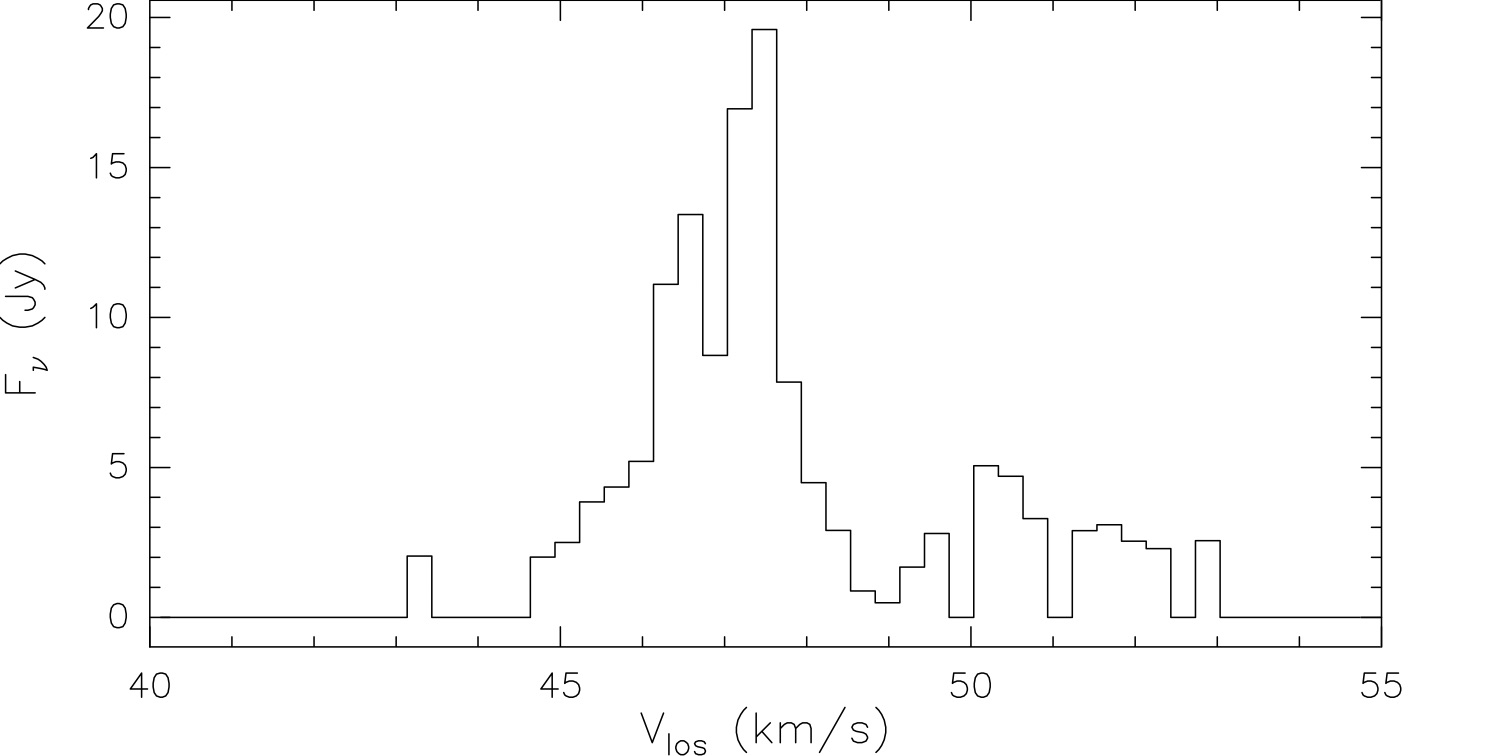}}
\caption{Upper-envelope spectrum for water-masers in o~Cet in $1990-2023$. The emission at velocities $>$49 \kms\ was only detected during two observations in December 2007 and March 2008.
}
\label{fig:ocet-upenv}
\end{figure}

\begin{figure}
\resizebox{9cm}{!}
{\includegraphics
{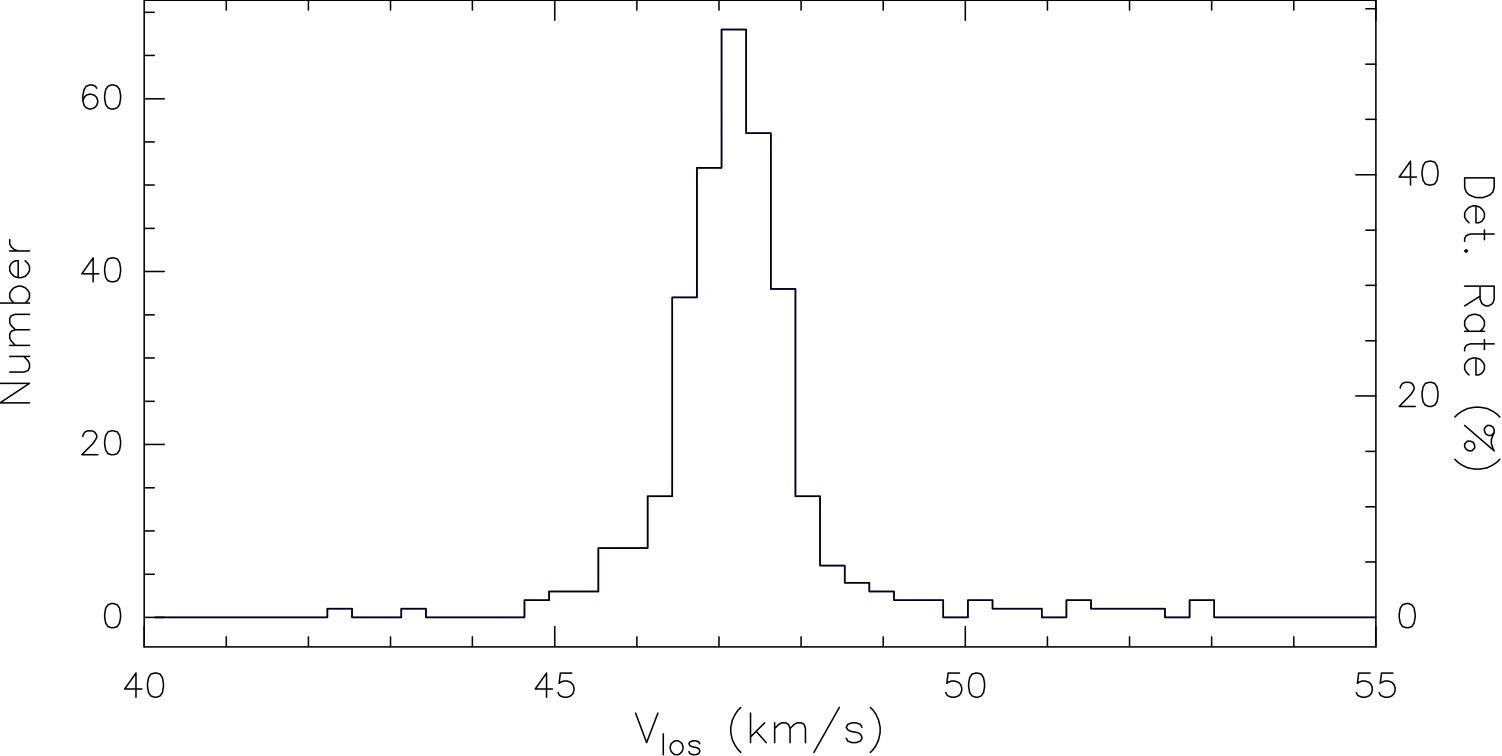}}
\caption{Detection-rate histogram for water-masers in o~Cet in $1990 - 2023$.
}
\label{fig:ocet-histo}
\end{figure}

The upper-envelope spectrum (Fig.~\ref{fig:ocet-upenv}) and the detection-rate histogram (Fig.~\ref{fig:ocet-histo} ) show that emission in general was detected between $45.6 < V_{\rm los} < 48.6$ \kms\ ($\Delta V_{\rm los} = 3.0$ \kms) for flux densities $> 3\sigma$ (Table \ref{centralcoords}). Like for R~Cas (Sect. \ref{profilevars}), we considered emission at velocities with fewer than 5 detections in the detection-rate histogram as spurious.

The narrowness and the asymmetry of $\Delta V_{\rm los}$ with respect to $V_{\ast}$ allows for only a marginal constraint on the outflow velocity $V_{\rm out}$ for the \water -maser region. The lower limit is $\Delta V_{\rm los}/2 < 2$ \kms, which is $<$25\% of the final expansion velocity of the CSE (Table~\ref{centralcoords}). The maser emission region could be located close to the star where the wind acceleration has started and move in the line of sight, or be moving in other directions and $\Delta V_{\rm los}/2$ gives the projected outflow velocity. It would then have to move at an absolute angle with respect to the line of sight $0^{\degr}  \le \theta < \sim76^{\degr}$. The lower-envelope is zero and not shown. 

The small \water -maser velocity range implies that the velocities of the maser lines making up the 47\,\kms\ feature could not have varied significantly. We made Gauss fits to all spectra with maser detections to derive the central velocity and the width of the 47\,\kms\ feature. Except for the single spectrum from autumn 1995 the profile could be fitted with a single line. 
Its mean velocity between 1995 and 2015 was $V_{\rm los} = 47.32\pm0.18$ and the mean width $FWHM = 1.10\pm0.38$ \kms. 
It is only the detection of two spatial components in late 1988 by \cite{bowers94} and the double-peaked spectral profile observed in the autumn of 1995 that indicate that the 47~\kms\ feature is likely a blend of at least two spatially separated emission components, which move over at least 21 years with almost constant velocity. The possible causes for the absence of traceable velocity variations has been discussed already for the case of the 24.5~\kms\ maser component of R\,Cas (Sect.~\ref{shellkinemat}). We consider it most likely that the movement of the 47~\kms\ feature is almost perpendicular to the line of sight.

\begin{figure}
\includegraphics[width=\columnwidth]
{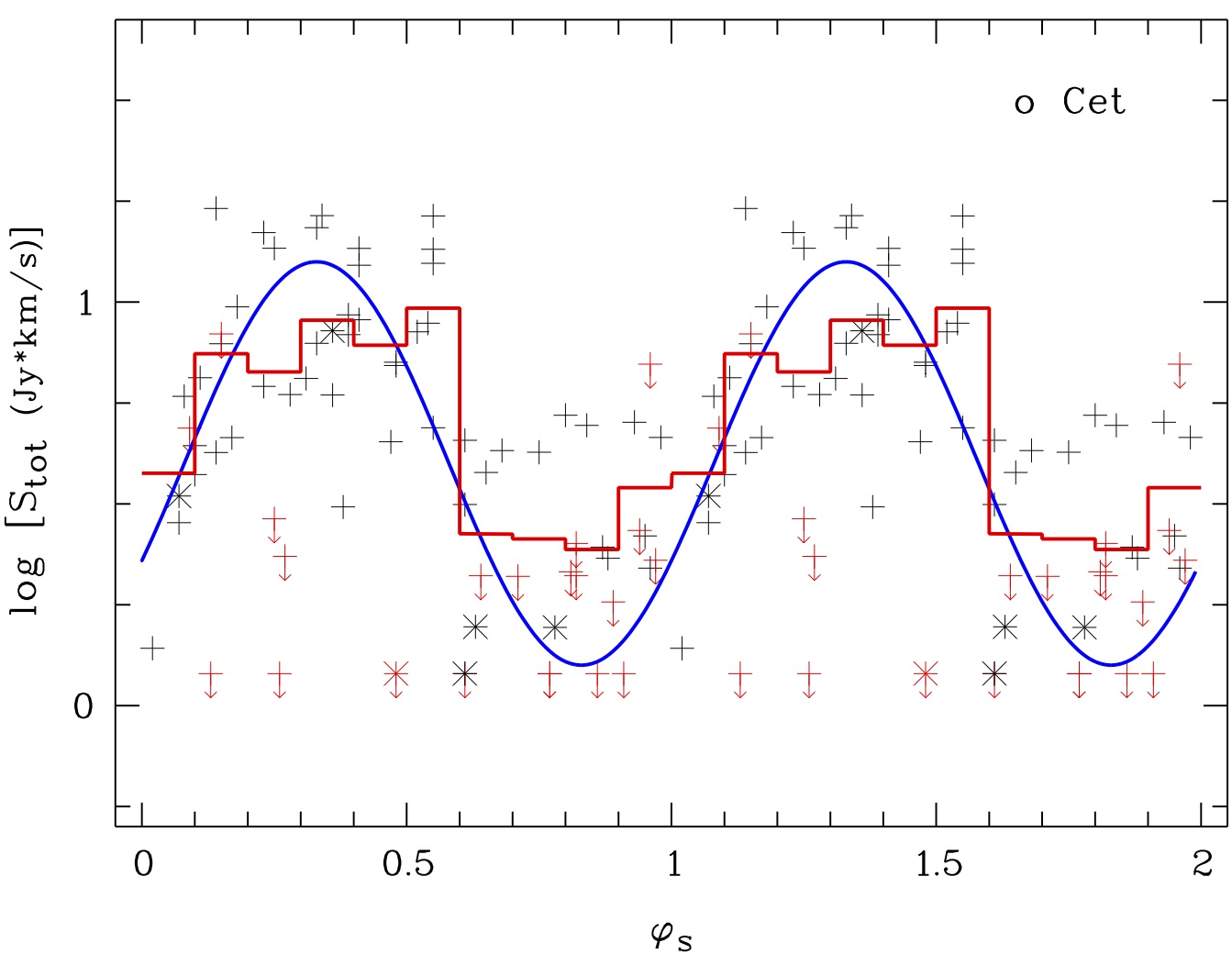}                
\caption{o~Cet \water -maser light curve in 1995 -- 2011. We plot the total fluxes $S_{\rm tot}$ vs. optical phase $\varphi_{\rm s}$. For better visualisation, the data are repeated for a second period and binned in phase intervals with a width of 0.1 (red step function). $\varphi_{\rm s}$=0 is defined as the time of maximum optical brightness. The data points marked by an asterisk are from Effelsberg, and the plusses are from Medicina data. Upper limits are shown as downward arrows in red. The sine curve (blue) was obtained by a fit to the $1995 - 2011$ radio measurements with a period $P_{\rm opt} = 332$ days, and it is delayed by $\phi_{\rm lag} = 0.33$, that is, by 110 days.  
}
\label{fig:ocet-lcurve}
\end{figure}

The \water -maser light curve of o~Cet shows periodic brightenings (see Fig. \ref{fig:ocet-fvt} and the light curve in \citealt{etoka17}). To analyse the maser light curve we determined integrated fluxes $S_{\rm tot}$ over the velocity range $+44 < V_{\rm los} < +50.5$ \kms. 
Because virtually all non-detections have $S_{\rm tot} < 1.2$~Jy\,\kms, the minimum upper limit in $S_{\rm tot}$ was set at 1.2~Jy\,\kms\ for non-detections, as long as $S_{\rm tot} \le 1.2$ Jy\,\kms \  was measured; otherwise the upper limit was set at the observed value. 

Using a Fourier analysis, we determined a period of $336\pm3$ days for the time range 1990 -- 2023, which is in accordance with the optical period and confirms that the brightenings are a response of the maser to the pulsations of the star. Using visual magnitudes from the AAVSO database taken between 1986 and end of 2023, we modelled the optical light curve by a sine-wave and related the maser light curve to this model  
(see Paper\,III for details). The optical phase $\varphi_{\rm s} = 0$ refers to the maximum of the optical model sine curve. To link the model to the observed light curve we use the date TJD$_{max}$ of the last optical maximum before 1.1.1987. The optical model light curve obtained has a period P$_{\rm opt} = 332$ days and a reference epoch  TJDmax = $6541 \pm 1$ days.

\begin{figure*}
\resizebox{18cm}{!}{
\includegraphics{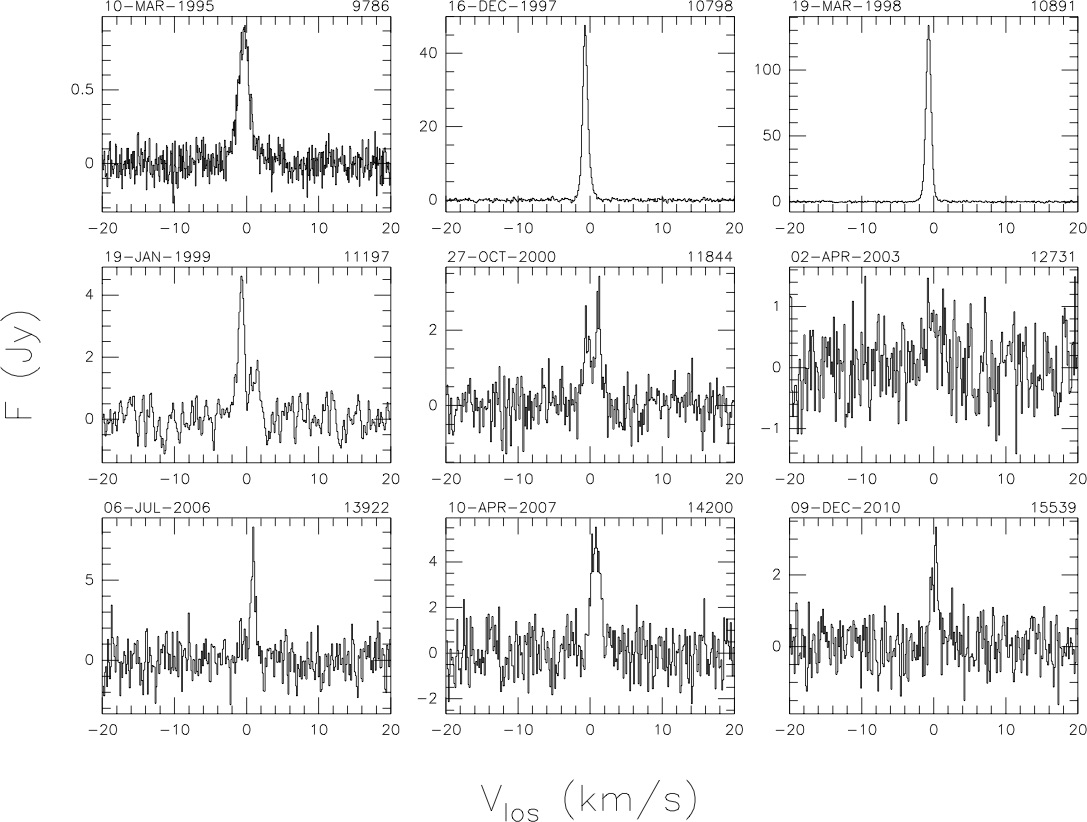}}
\caption{Selected H$_2$O maser spectra of R~Leo. The calendar date of the observation is indicated in the top left corner above each panel, the TJD (JD-2440000.5), in the top right corner. 
}
\label{fig:rleo_sel}
\end{figure*}

Due to the small number of observations in the years $<$1995, and 2012 -- 2017, and the small number of maser detections after 2017, we restricted the analysis of the maser light curve to the years 1995 -- 2011.  Fig. \ref{fig:ocet-lcurve} shows the maser light curve using observations in this time interval relative to the phase $\varphi_{\rm s}$ of the optical light curve. The radio emission for a particular phase shows a large scatter and the lag of the radio light curve is $\phi_{\rm lag} = 0.33$ (Table \ref{centralcoords}). Consistent with this lag of the maser light curve relative to the optical one, the strongest maser peaks in the 2018 -- 2023 observing period were seen in April/May 2019 ($\varphi_{\rm s} = 0.27-0.34$) and in Jan./Feb. 2021 ($\varphi_{\rm s} = 0.19-0.29$). The \water -maser phase lag of o~Cet is similar to the one observed for R\,Cas (Table \ref{centralcoords}) and of the same order as those of the Mira variables U\,Her and RR\,Aql. As discussed in Paper\,III and following \cite{smith06}, we attribute the general presence of this lag to the presence of strong titanium oxide (TiO) absorptions in the visual band  of O-rich Mira variables, which curtails the rise in the optical light and makes the visual maximum precede the actual maximum of stellar luminosity, and hence of the \water -maser emission by 20 -- 30\% of the stellar period. 

The \water -maser emission from o~Cet comes from a region with a radius of a few au \citep{lane87, bowers94}, which is small compared to the typical \water -maser shell radii ($6-45$~au)  observed for other SR- and Mira variables (\citealt{bains03}; Paper I -- III; this paper). It is instructive to discuss the \water -maser velocity range and the size of the emission region in relation to the SiO-maser emission in this star. The v=1 J=1--0 SiO-maser emission ($\nu = 43.12$~GHz) is spread over a velocity range $43.5 - 54.5$~\kms\ \citep{jewell91}, which is asymmetric with respect to the stellar velocity and much broader than the \water -maser velocity range. The SiO emission has been mapped in several epochs and is located in a ring with an average diameter of 70 mas \citep{cotton08}, i.e. 6.4 au. The ring is located inside the inner boundary of the dust shell, having a diameter of 11~au as measured at 11~\micron\ \citep{danchi94}.

The small size of the observed \water -maser region is therefore likely not the size of an \water -maser shell, but only of a region within the stellar wind of the star. The cause could be the lack of \water\ molecules at those (larger) radial distances, where \water -maser emission normally emerges. \cite{etoka17} have observed OH main-line bursts at projected distances of less than $0.4\pm0.04$\arcsec\ ($\sim40\pm4$ au) from the star, and infer that enhanced OH production by photodissociation of \water\ occurs at these distances. The \water -masers of o~Cet may therefore come from those parts of the stellar wind, at smaller distances from the star, where \water\ molecules have survived.

The reason for the enhanced OH production is unknown but could be linked to the presence of the companion Mira B emitting UV radiation \citep{reimers85} and having a separation in 2014 of 0.472\arcsec\ (43.4 au) from o~Cet (Mira A) \citep{vlemmings15}. The orbital movement of Mira A around the barycenter of the Mira AB-system should induce an acceleration of the systemic velocity, which was estimated by \cite{etoka17} to be about 0.05 \kms\ yr$^{-1}$. This is not seen in our data, however, because from the profile fitting of the single dominant 47\,\kms\ maser feature an upper limit for a velocity drift of 0.01 \kms\ yr$^{-1}$ can be derived. We conclude that for the years 1995 -- 2015 a change in the radial velocity of o~Cet as large as considered by \cite{etoka17} could not be confirmed.

The outflow velocity of the region is not well constrained, but assuming a lower limit of $V_{\rm out} = 2$ \kms, the region (considered as a density-bound cloud) between 1995 and 2015 would have moved over a distance of $>8$~au. Even with a modest velocity gradient $K_{\rm grad} = 0.1$ \kms\ au$^{-1}$ \citep{richards12} a shift of $>0.8$ \kms\ should have been observed, if the region would have survived and would have moved in the line of sight. As this is not observed, the region is either moving in directions almost perpendicular to the line of sight or the region is stationary and the \water\ molecules are excited and emit maser emission as soon as they enter the region. As the maser clouds in Mira variables (and SRV) are considered to have lifetimes of at most a few years (\citealt{bains03}; Paper I and III), we consider a stationary \water -maser region in the wind of o~Cet more likely. 

\cite{etoka17} did not find a unique position of the OH-maser flaring regions relative to the Mira AB axis, inhibiting a prediction for the position of the \water -maser region relative to this axis. Moreover, the small velocity offset of $\sim+0.8$ \kms\ between the 47 \kms\ feature and the adopted radial velocity of o~Cet (see Fig. \ref{fig:ocet-fvt}) is too small to reliably infer a location of the \water -maser on the rear side of the star, given the relatively high uncertainty of the stellar radial velocity.

To summarise, the \water -maser of o~Cet is likely coming from a region within the stellar wind, which is small compared to the usual \water -maser shell sizes in Mira variables. The absence of a velocity drift over almost $\sim$20 years constrains the acceleration due to the orbital motion of the binary system in the years 1995 -- 2015 to less than 0.01 \kms\ yr$^{-1}$. We infer that the \water -maser is not hosted by a density bounded cloud moving with the wind but is most likely excited in a stationary region. The maser brightness variations are linked to the stellar optical light variations, but are delayed by about one third of the period.

\section{R~Leo \label{sec:rleo}}

R~Leo is a Mira variable with a period of 310 days. It is the closest star in our sample with a distance of $71^{+17}_{-11}$ pc. We adopt a stellar radial velocity $V_{\ast} = 0.0$ and a final expansion velocity $V_{\rm exp} = 9.0$ \kms, as determined from circumstellar CO and thermal SiO emission (see Table~\ref{centralcoords}). 

The \water -maser of R~Leo was first detected in 1973 by \cite{dickinson76} as a single feature ($F_\nu \sim 8$ Jy) at $V_{\rm los} = -1$ \kms. Subsequent observations showed strong variability with non-detections as well as detections with flux densities reaching several hundred Jy. This triggered extensive monitoring observations with the Pushchino radio telescope over twenty years 1980 -- 1999  (\citealt{esipov99} and references therein). A spectacular emission burst reaching $\sim740$ Jy and lasting about half a year was seen with Pushchino in 1982 \citep{rudnitskij87}. In October 1983, in the aftermath of the burst, the maser still had a flux density of $\sim$20\,Jy and was mapped by \cite{lane87}, who reported a diameter of the emission region of $\le7$ au. Subsequently the maser was  not detected ($F_\nu < 10$ Jy; 3$\sigma$) until 1997 \citep{esipov99}. During this time  non-detections were reported in 1991/92 by \cite{takaba94} and \cite{takaba01}, and in 1995 by \cite{imai97a} using VLBI. Beginning in 1997 \citeauthor{esipov99} observed a second burst with a maximum peak flux density of 140 Jy observed in March 1998. The maser brightness then declined rapidly and was not detected anymore until the end of their monitoring programme in January 1999.

We started our observations in 1988 and performed a monitoring programme with regular revisits in 1995 -- 2011. Some additional spectra were taken in 2015. In Fig.~\ref{fig:rleo_sel} we show selected maser spectra of this star. The complete set of 99 spectra is shown in Fig.\,A.3 (Appendix, on \href{https://zenodo.org/records/15534987}{Zenodo}). 

\begin{figure}
\includegraphics[width=\columnwidth]
{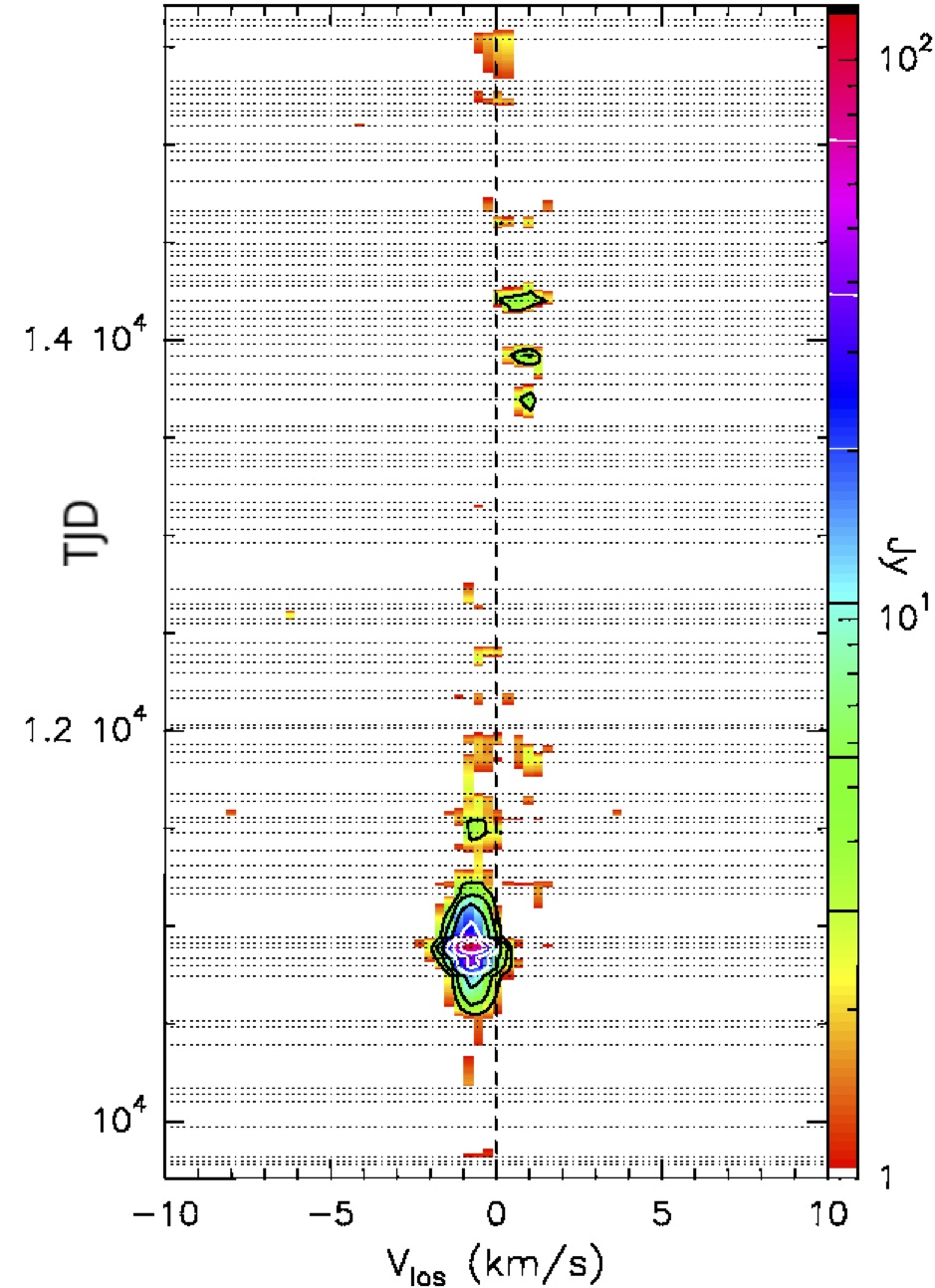}
\caption{As Fig.~\ref{fig:rcas-fvt}, but for R~Leo. The first spectrum was taken on
10 March 1995 (JD = 2449786.5, TJD = 9786), and the last spectrum was taken on 20 March 2011.}
\label{fig:rleo-fvt}
\end{figure}

\begin{figure}[!b]
\includegraphics[width=\columnwidth]
{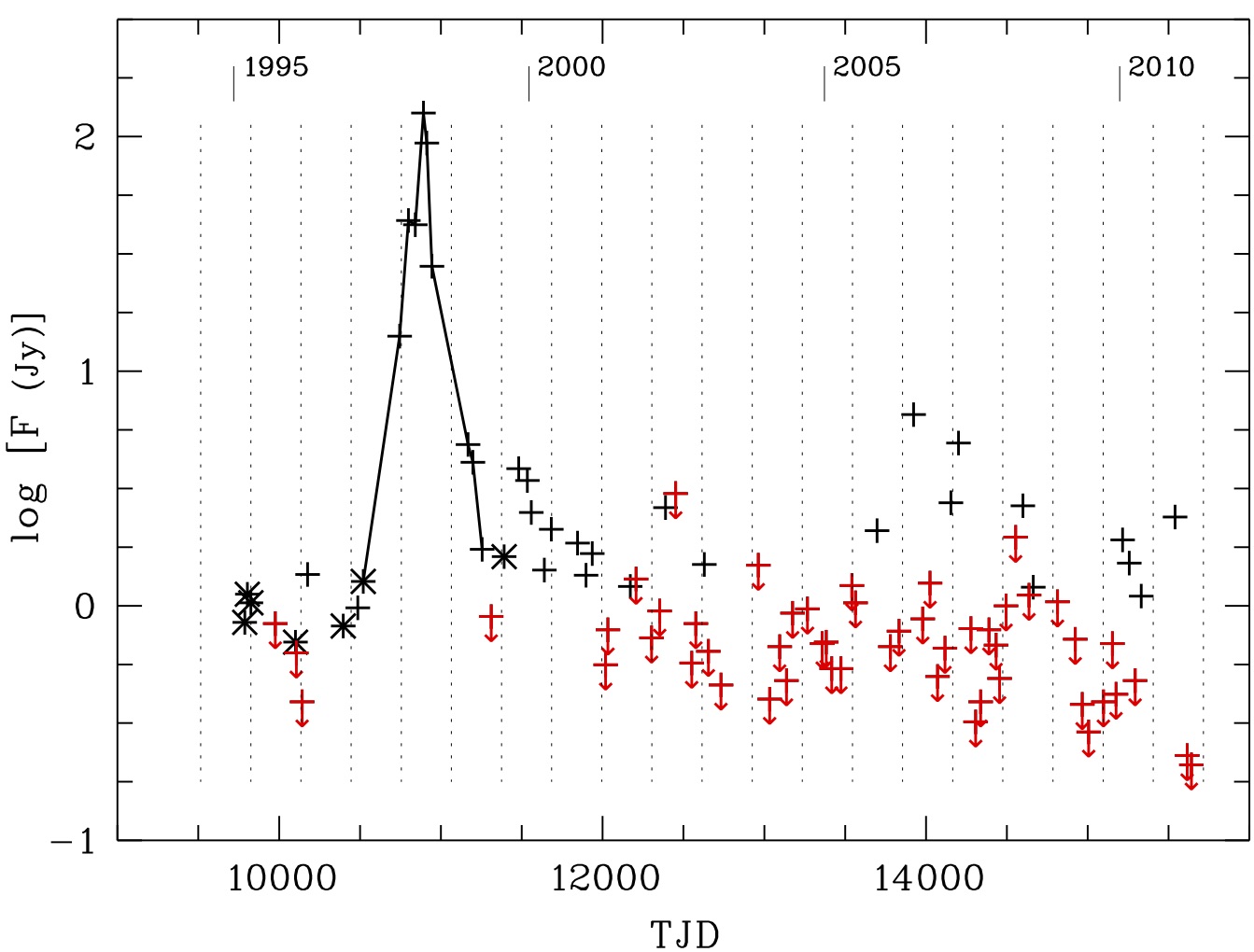}                
\caption{R~Leo \water -maser light curve in 1995 -- 2011. We display the flux densities obtained from the line fit features seen between $-2$ and $+2$ \kms\ (see text) vs. TJD. Data points marked by an asterisk are from Effelsberg. Upper limits are shown as downward arrows in red. Times of maximum optical brightness defined by the sine wave fit to the optical data are indicated by dotted lines. Observations during the burst in 1997 -- 1999 (TJD $\sim$ 11000) are connected by solid lines. 
}
\label{fig:rleo-lcurve-two}
\end{figure}

Our first Medicina observation, in February 1988 (Fig.\,A.3; Appendix, on \href{https://zenodo.org/records/15534987}{Zenodo}), resulted in a non-detection, and was already reported by \cite{comoretto90}. During our observations in 1990 and 1994, the maser was not detected either or was weak. As shown in the FVt-plot (Fig.~\ref{fig:rleo-fvt}), the maser was weak on the $<2$ Jy level for the brightest peak during our complete monitoring programme 1995 -- 2011, except for a burst interval of about 18 months between October 1997 and January 1999 (TJD = 10700 -- 11200), also observed by \cite{esipov99} in its brightest part. With very few exceptions, the maser was detected with Medicina only marginally ($\approx1\sigma$; Medicina spectra have an rms $\approx$ 0.5 -- 1.0 Jy) outside the burst period, whereas the Effelsberg spectra (1995 -- 1999; rms = 0.1 -- 0.25 Jy) always showed clear detections of a single feature on the 1.0 Jy level. The light curve (Fig.~\ref{fig:rleo-lcurve-two}) is therefore ill-defined outside the burst period. Most of the spectra show evidence for emission, however, suggesting that the maser was never really extinguished during the years of monitoring. In June 2009 \cite{kim10} failed to detect the maser, which matches our non-detection in the same month. In 2015 the maser was always detected by us, but stayed at $<3$ Jy in flux density. 

\begin{figure}
\includegraphics[width=\columnwidth]
{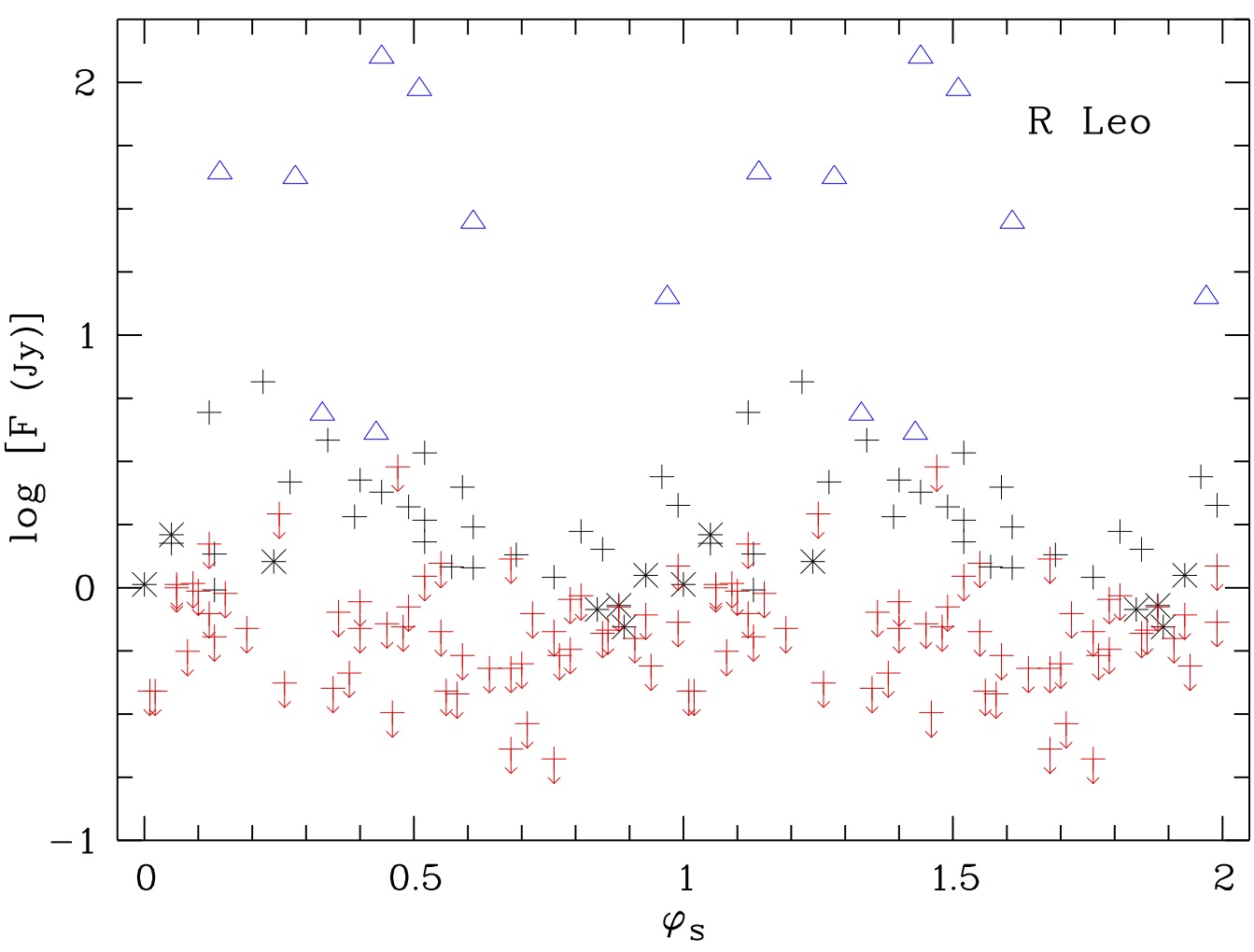}
\caption{ R~Leo \water -maser light curve in 1995 -- 2011. We plot the peak flux densities (from the line fit) vs. optical phase $\varphi_s$. See Fig. \ref{fig:ocet-lcurve} for further details. The blue triangles belong to the burst period (see Fig. \ref{fig:rleo-lcurve-two}).
}
\label{fig:rleo-lcurve}
\end{figure}

\begin{figure}
\resizebox{9cm}{!}
{\includegraphics
{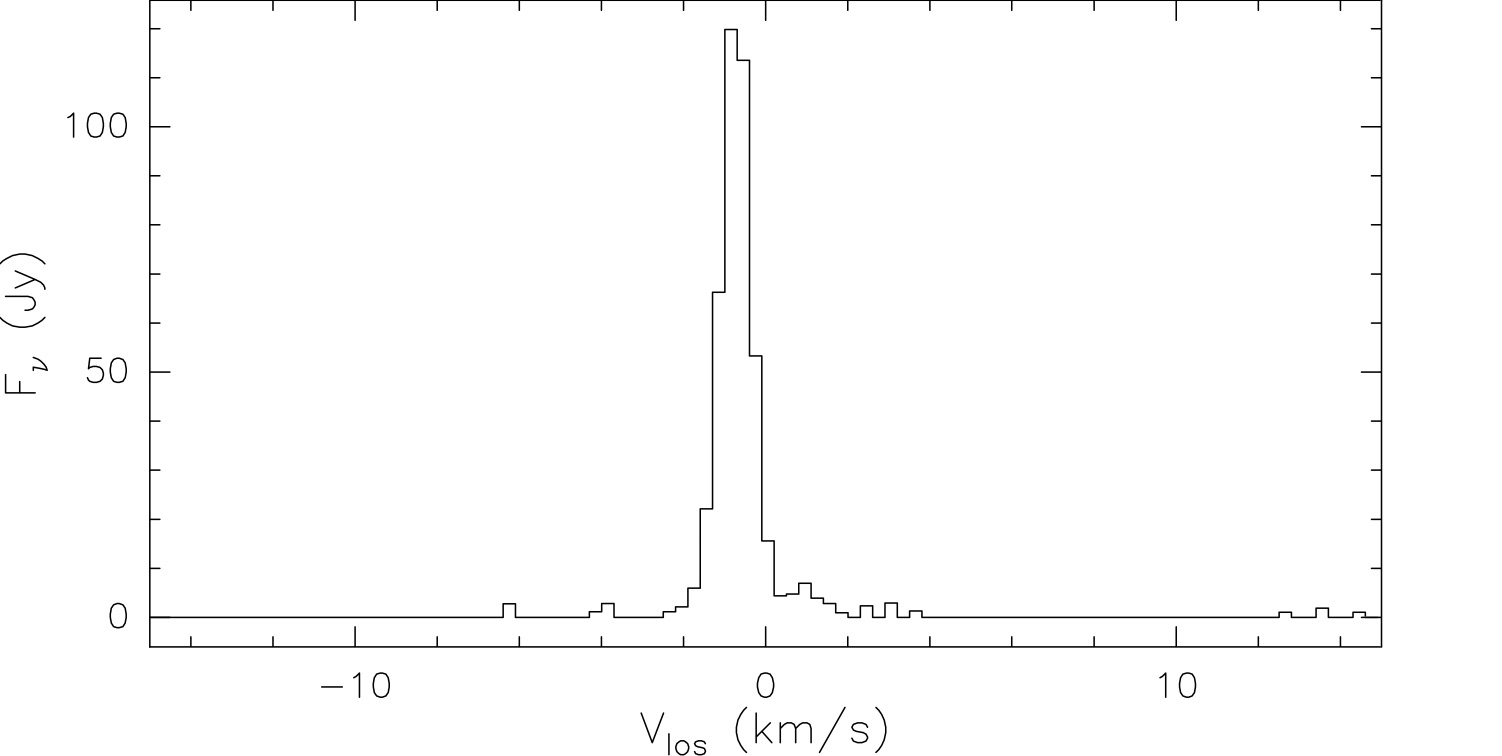}}
\caption{Upper-envelope spectrum for water-masers in R~Leo in $1988 - 2015$. 
}
\label{fig:rleo-upenv}
\end{figure}

\begin{figure}
\resizebox{9cm}{!}
{\includegraphics
{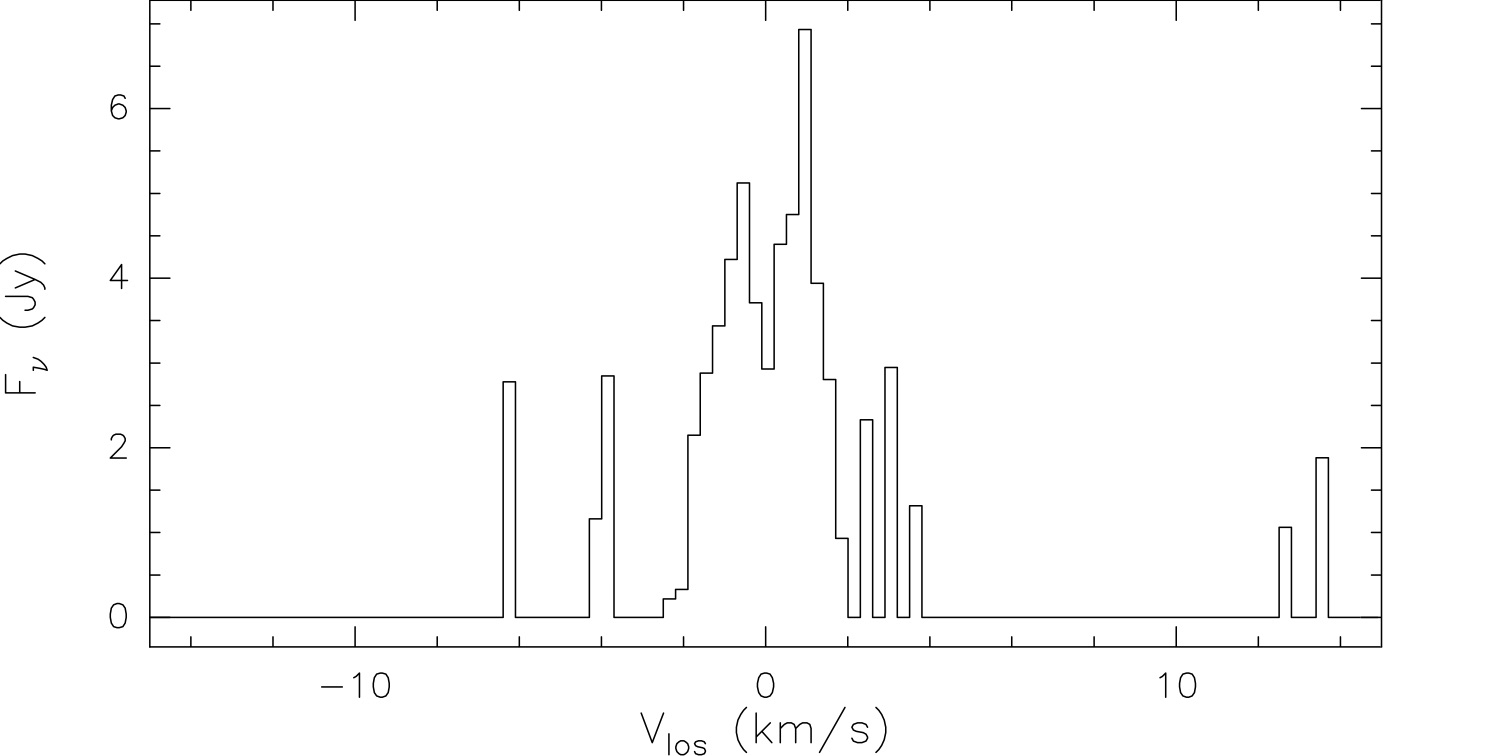}}
\caption{As Fig.~\ref{fig:rleo-upenv}, but without the six spectra that were involved in the burst (TJD $10744-10944$: 23 October 1997 to 11 May 1998). 
}
\label{fig:rleo-upenv-noflares}
\end{figure}

\begin{figure}
\resizebox{9cm}{!}
{\includegraphics
{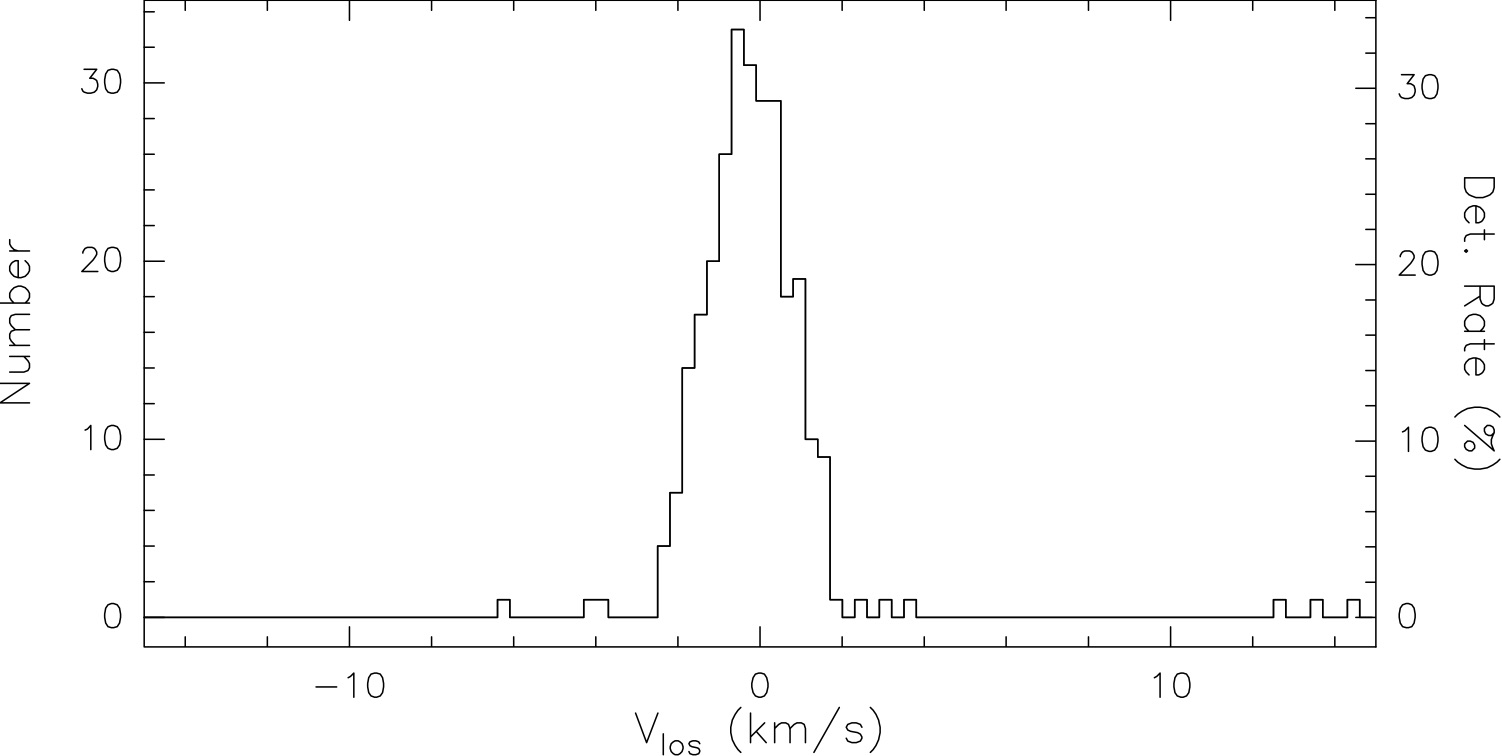}}
\caption{Detection-rate histogram for water-masers in R~Leo in $1988 - 2015$. 
}
\label{fig:rleo-histo}
\end{figure}

The velocity boundaries for detected maser emission $1988 - 2011$ were determined as for R~Cas and o~Cet (Sect. \ref{profilevars} and \ref{sec.ocet}, respectively). The upper-envelope spectrum (Figs.~\ref{fig:rleo-upenv} and \ref{fig:rleo-upenv-noflares}) and the detection-rate histogram (Fig.~\ref{fig:rleo-histo} ) show that emission with flux densities $> 3\sigma$ was detected in the range $-2.2 < V_{\rm los} < 1.7$ \kms\ ($\Delta V_{\rm los} = 3.9$ \kms) (Table \ref{centralcoords}). The lower-envelope is zero and not shown. The projected outflow velocity of the \water -maser region of $\Delta V_{\rm los}/2$ $\sim 2$ \kms\ is rather small and would reach less than 25\% of the final expansion velocity. 

This velocity difference is similar to the case of o~Cet (Sect.~\ref{sec.ocet}), where in addition the SiO-maser emission is spread over a larger velocity range than the \water\  maser emission, although the SiO emission originates at smaller radial distances than the \water\  emission. For R~Leo, the v=1 J = 1--0 SiO-maser emission is also spread over a much wider velocity range. In 1983/1984 SiO emission was seen for R~Leo at $-9$ to $+5$ \kms\ \citep{jewell91}, while in 2009 the velocity range with emission was $-3$ to $+10$ \kms\ \citep{desmurs14}. In analogy to the \water -maser emission region in o~Cet, in R~Leo the observed maser emission may trace only a part of the water-maser shell.

To construct a maser light curve we made a Gaussian fit to all spectra, assuming the presence of a single maser feature in the velocity range $-2$ to $+2$ \kms.  For features with flux densities $<1.5$ Jy the fit is usually poor in Medicina spectra and we treat the noise of the spectra ($1\sigma$) rebinned to a channel width of 0.3 \kms\ as upper limits. 
This is justified because in many Medicina spectra the presence of water-maser emission is clearly seen at the level of the noise. There are only few $>1\sigma$ detections outside the burst period. The light curve based on the peak flux density of the line fit is shown in Fig. \ref{fig:rleo-lcurve-two}. This light curve basically confirms the properties of the maser as observed at Pushchino since 1984 in that the flux density is very low for most of the time and increases only occasionally by factors of up to two magnitudes within bursts lasting $\sim$1 year. To explain the initially observed strong variability of the \water -maser of R\,Leo between its discovery and 1983, another burst may have occurred in late 1976 or the maser was on average brighter (see the discussion in \citealt{esipov99}).

The optical phase $\varphi_{\rm s}$ of the stellar light variation was determined as in the case for o~Cet, with  $\varphi_{\rm s} = 0$ defined by the sine function fitted to the optical light curve (1987 -- 2015) with a period of $P_{\rm opt} = 310$ days, and using TJD$_{max}$=6725. The radio detections were too sparse for us to determine the periodicity. As function of the optical phase, the \water -maser observations of R~Leo are shown in Fig. \ref{fig:rleo-lcurve}. The data points belonging to the burst period are singled out. The burst lasted significantly longer ($\sim1.5 P_{\rm opt}$) than the duration of the pulsation period and does not seem to be related to the optical light curve. As shown in Fig. \ref{fig:rleo-lcurve-two}, the burst reached its maximum close to minimum light. Consistent with the results of \cite{esipov99}, the flux density reached $\approx130$ Jy, a factor of 50 -- 100 higher than the 'quiescent' level. From the few clear line detections outside the burst period the two strongest occurred at phases $0 < \varphi_{\rm s} < 0.3$ in agreement with expectations, if the water-maser and optical emissions are correlated with a phase-lag. Thus, the \water -maser emission in general may behave as in other Miras, but was generally too weak to be adequately detected. The 1998 burst, as well as the one in 1982, obviously was an event not related to the periodic optical variations.

During the burst in 1997 -- 1999 the velocity of the maser feature was $\sim-0.6$ \kms. The velocities of the four Medicina detections in 2005 -- 2011 (TJD=13500 -- 16000; see Fig. \ref{fig:rleo-lcurve-two}) were between +0.1 and +1.0 \kms, showing that different emission regions were active during the monitoring years. The three spectra taken between October 2000 and January 2001 (TJD$\approx11900$) (Fig.~\ref{fig:rleo_sel} and Fig.\,A.3 on \href{https://zenodo.org/records/15534987}{Zenodo}) show evidence that there were two maser features of equally weak flux density present in the velocity interval $-1 < V_{\rm los} < +1$ \kms.

To summarise, the \water -maser properties of R~Leo appear to be similar to those of o~Cet, although on average the emission is weaker. The difference is the occurrence of strong bursts (likely 1976, 1982, and 1998) over $\sim20$ years before 2000, but not in the decade thereafter. The size of the emission region imaged in 1983 is also of the same order as in o~Cet. It is possible that in R~Leo at a given time only particular regions within the \water -maser shell are detected, in which the brightness of the otherwise regularly present maser emission surpassed available sensitivity limits.

\section{$\chi$~Cyg \label{sec:khicyg}}
$\chi$~Cyg is an oxygen-rich Mira variable of spectral type S7 -- S10 (\citealt{keenan74}). It has been detected as SiO-maser source in several transitions (\citealt{cho07} and references therein), while there were several attempts to detect an \water -maser (\citealt{dickinson76}; \citealt{bowers84}; \citealt{engels88}; \citealt{takaba94}) but without success. With a distance of 160 pc (Table \ref{centralcoords}) $\chi$~Cyg is one of the closer Mira variables, making the failures peculiar. We monitored the star regularly to test if the non-detections in the past were due to variability or due to real absence. During $\sim$15 years (1995 -- 2009) 72 observations were made without a single detection. This observation is unique, because never a star without previously known \water -maser emission had been monitored with such a high cadence ($\approx$ every 3 months) before, covering several periods. The non-detections are corroborated by non-detections in $\sim2003$ by \cite{shintani08} and in 2009 by \cite{cho12}.

\begin{figure}
\includegraphics[width=\columnwidth]
{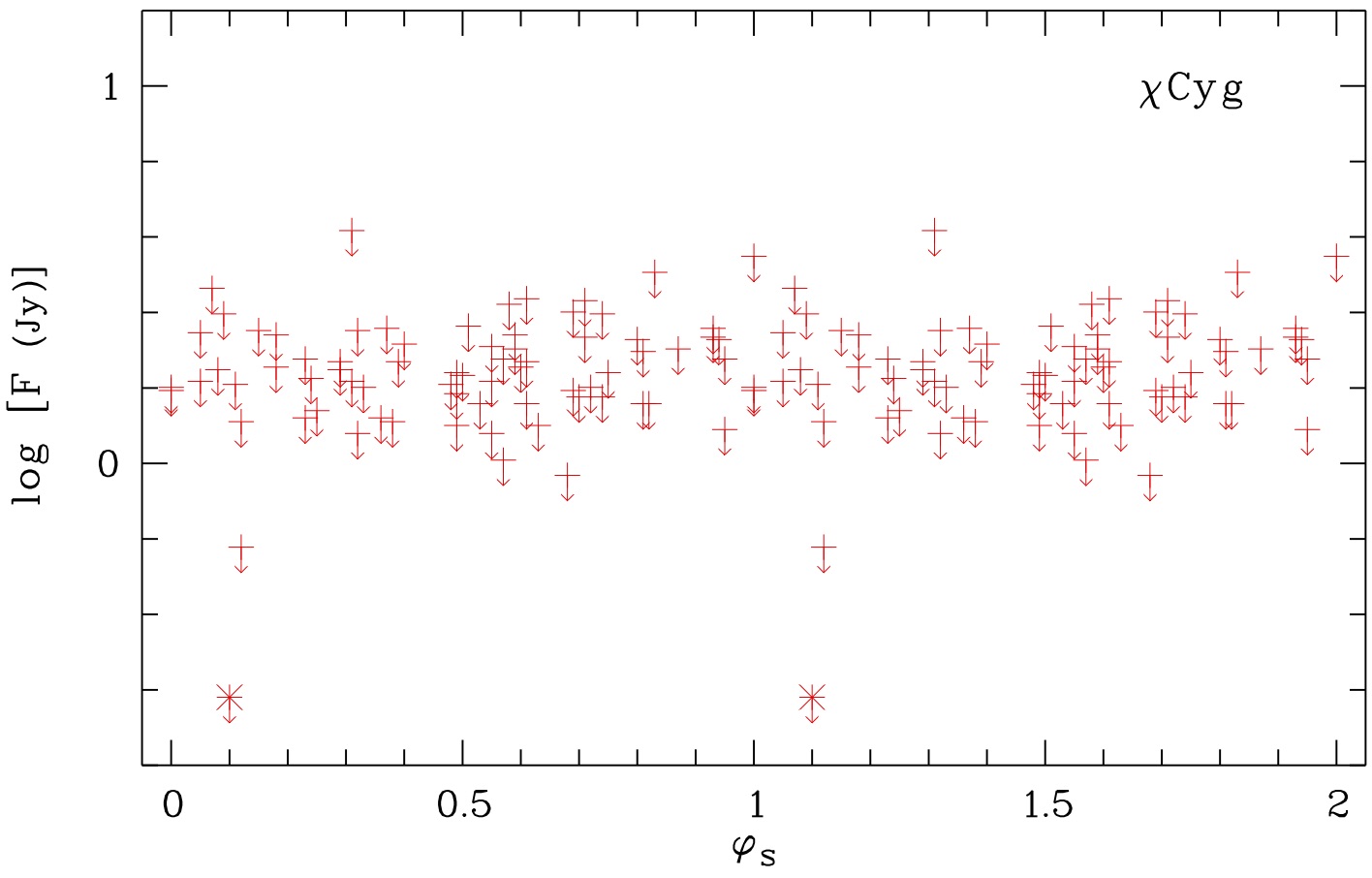}
\caption{ $\chi$~Cyg \water -maser observations between September 1995 and June 2009. The data point marked by an asterisk is from Effelsberg, the plusses are Medicina data. We plot the 3$\sigma$ flux density upper limits, where $\sigma$ is the noise of the spectra. There are no detections. 
}
\label{fig:khicyg-lcurve}
\end{figure}

$\chi$~Cyg  has an optical period $P_{\rm opt} = 407$ days, so that the monitoring programme covered more than 13 cycles. The optical phase $\varphi_s$ was determined as for o~Cet with a fit of a sine wave to the optical data. We used three times the noise of the maser spectra rebinned to a channel width of 0.3 \kms\ (rms $\approx0.6$ Jy for Medicina) as upper limit for maser emission in the velocity interval $-30 < V_{\rm los} < 50$ \kms. 
The radio 'light curve' showing these upper limits as function of the optical phase is shown in Fig.\ref{fig:khicyg-lcurve}. All optical phases are well covered. 

The persistent non-detection of \water -maser emission in $\chi$~Cyg supports the general absence of \water -masers in S stars \citep{jorissen98}. If they exist, they may have such low luminosities that they rarely have flux densities surpassing typical sensitivity limits of past \water -maser surveys (rms $\sim 0.1-1$\,Jy), except in individual cases as in the S star R\,Cyg \citep{takaba01}. The absence of the \water -masers may be explained by their chemistry. Having a carbon-to-oxygen ratio C/O$\approx$1, free oxygen may be used up for molecules like SiO in the inner shell, inhibiting the formation of \water\ and also OH at larger radial distances from the star. \cite{danilovich14} find for the S star W\,Aql that the \water\ abundance is indeed decreased compared to M stars, which are the typical hosts of \water\ and OH-masers.

\section{Discussion}

The long-term monitoring of the \water -masers reveals that in addition to properties they have in common, the brightness variations in the emission in all four stars show unique behaviour. This applies as well to the \water -maser properties of SRVs and Mira variables, which we have analysed in Papers I -- III.
Here we focus on a comparison of the maser luminosities, the outflow velocities in the \water -maser shell, and on lessons learned with regard to the use of \water -masers to study the structure and evolution of the stellar wind in the CSE at  distances of a few tens of au from the star. 

\subsection{\water -maser luminosities \label{maserlumo}}

\begin{table*}
\caption{\water -maser luminosities, stellar luminosities, and mass-loss rates of the stars we studied so far. }
\label{table:photon-luminosities}
\begin{center}
\begin{tabular}{lrr|c|rr|rr|rr|c|c}
\hline\noalign{\smallskip}
\multicolumn{1}{c}{Star} & \multicolumn{1}{c}{Type} & \multicolumn{1}{c|}{$D$} & log\,$L_{\rm H_2O}^{\rm up}$ &
\multicolumn{6}{c|}{log\,$S_{\rm tot}$, log\,$L_{\rm p}$} & log\,$L_{\rm bol}$ & log\,\mdot \\[0.05cm]
\multicolumn{2}{c}{} & \multicolumn{1}{c|}{[pc]} & [L$_{\odot}$] &
\multicolumn{6}{c|}{[Jy~\kms] , [s$^{-1}$]} & [L$_{\odot}$]    &  [\Myr]   \\
&&&& \multicolumn{2}{c|}{High} & \multicolumn{2}{c|}{Mean}     & \multicolumn{2}{c|}{Low} &&\\
\hline\noalign{\smallskip}
R~Crt    & SRV &  236 & $-$4.88& 3.5 & 44.0& 2.9 & 43.4 &    2.3 &    42.8 &  $4.03\pm0.10$ & $-5.46$ \\
RT~Vir   & SRV &  226 & $-$5.16& 3.3 & 43.8& 2.8 & 43.3 &    2.3 &        42.8 &  $3.70\pm0.09$ & $-6.05$ \\
RX~Boo   & SRV &  136 & $-$6.23& 2.6 & 42.7& 1.9 & 41.9 &    1.5 &    41.5 &  $3.58\pm0.11$ & $-6.12$ \\
SV~Peg   & SRV &  333 & $-$6.42& 1.6 & 42.4& 0.9 & 41.7 & $<$0.8 & $<$41.6 &  $3.93\pm0.20$ & $-6.04$ \\
o~Cet    &Mira &   92 & $-$8.16& 1.2 & 40.9& 0.5 & 40.2 & $<$0.08 & $<$39.8 & $3.54\pm0.16$ & $-$6.73 \\
R~Leo    &Mira &   71 & $-$7.83$^{\dagger}$& 1.7 & 41.1& $<$0.7 & $<$40.2 & $-$ & $-$ & $3.23\pm0.15$ & $-$7.16 \\
U~Her    &Mira &  266 & $-$5.77& 2.6 & 43.3& 2.0 & 42.6 &    1.5 &    42.1 &  $3.71\pm0.17$  & $-$6.25 \\
$\chi$ Cyg&Mira &  160 & \tB{$-$} & $<$0.7 & $<$40.9 & $-$ & $-$  & $-$ & $-$ & $3.79\pm0.10$ & $-$6.56 \\
RR~Aql   &Mira &  410 & $-$5.37& 2.7 & 43.9& 2.1 & 43.3 &    1.4 &    42.6& $3.75\pm0.15$ & $-$5.84   \\
R~Cas    &Mira &  174 & $-$6.14$^{\dagger}$& 2.3 & 42.6 & 1.0 & 41.2 & $<$0.7 & $<$41.0 & $3.75\pm0.15$ & $-5.97$ \\
\noalign{\smallskip}\hline
\end{tabular}
\end{center}
Notes: The stellar type is semi-regular variable (SRV) or Mira.
We list distances $D$ with their references as noted below.
Characteristic levels of \water -maser brightness ($S_{\rm tot}$ = integrated flux densities) and maser luminosities (\Lup  (L$_{\odot}$); $L_{\rm p}$(photons s$^{-1}$)) are listed for the objects from Paper~I in 1987 -- 2005, from Papers II and III in 1987 -- 2015, and from this paper in $1988 - 2015$ or $1987/1990 - 2023$. The definition of the levels (high, mean, and low) is described in the main text. Columns $\log$\,$L_{\rm bol}$ and $\log$\,\mdot\ list stellar luminosities and mass-loss rates, respectively.\\
Reference for distances: see Table~\ref{centralcoords}; Papers I, II and III. \\
$^{\dagger}$: The \water -maser luminosity derived from the upper-envelope spectrum \Lup\ excluding the burst periods are log\,$L_{\rm H_2O}^{\rm up} = -8.69$ \Lsun\ for R\,Leo (6 observations ignored) and log\,$L_{\rm H_2O}^{\rm up} = -6.90$ \Lsun\ for R\,Cas (7 observations ignored). 
\end{table*}

In Table~\ref{table:photon-luminosities}, we give information on luminosity and other parameters of the four Mira variables presented in this paper, as well as those of the SRVs (R\,Crt, RT\,Vir, RX\,Boo, SV\,Peg) and Mira variables (U\,Her, RR\,Aql) treated in Paper I, II, and III. For the details of how the luminosities were derived, we refer to Paper~III. \Lup, the potential maximum \water -maser luminosity derived from the upper-envelope spectrum, represents the maximum output which the source could produce if all the velocity components would emit at their maximum level, at the same time and equally in all directions. Characteristic levels (high, mean, and low) of maser brightnesses are listed as well, as given by integrated flux densities $S_{\rm tot}$ in Jy \kms\ and corresponding maser luminosities $L_{\rm p}$ in photons per second. The brightness of the mean level is the median of all integrated flux density measurements, 
while the high and low levels are represented by the median of the seven highest and lowest integrated flux density measurements, respectively. To determine the maser luminosities for R\,Leo, we obtained integrated flux densities in the velocity range $-5 \le V_{\rm los} \le 5$ \kms\ and adopted a minimum upper limit of $S_{\rm tot}$ = 5 Jy\,\kms\ for non-detections with measured $S_{\rm tot}$ below this limit. The same minimum upper limit for integrated flux densities was used for $\chi$\,Cyg.

The mean \water -maser luminosities of o\,Cet and R\,Leo ($\log L_{\rm p} [{\rm s}^{-1}] \le 40.2$) are the lowest among the Mira variables and SRVs monitored. Most observations did not detect the \water -maser of R\,Leo and therefore only an upper limit for the mean-luminosity level can be given, while the low-level luminosity is not available. It was only during the burst phase 1997/98 that R\,Leo's maser luminosity reached an appreciable level ($\log L_{\rm p} [{\rm s}^{-1}] \sim 41$), which is still lower than the mean-luminosity level of the other stars observed, except for o\,Cet. During the maximum of the burst the maser luminosity was at least a factor of $8$ higher than the mean luminosity of R Leo. For o\,Cet, a star without burst during our monitoring programme, the ratio of the high and mean level is a factor of 5, which can be attributed entirely to the maser luminosity variations caused by the stellar pulsation (Fig. \ref{fig:ocet-lcurve}). 
 
The mean-luminosity level of R\,Cas ($\log L_{\rm p} [{\rm s}^{-1}] = 41.2$) is intermediate between o\,Cet/R\,Leo and the prominent \water -maser emitters R\,Crt, RT\,Vir, U\,Her and RR\,Aql, and is similar to the weaker SRVs RX\,Boo and SV\,Peg. Thus, Mira variables and SRVs emit \water -maser emission with a range of mean luminosities spread by a factor of at least 1000. The upper limits for the low luminosity levels in Table \ref{table:photon-luminosities} indicate that \water -masers of stars with distances $\la 400$\,pc from the Sun needed to surpass luminosities $\log L_{\rm p} [{\rm s}^{-1}] > \sim40$ for having been detected by early \water -maser surveys with sensitivity limits at the Jansky level from which the current sample monitored was selected. 

The upper limits to the maser luminosity levels given in Table \ref{table:photon-luminosities} are certainly determined by our sensitivity limits, which are dominated by the Medicina observations, that greatly outnumber those made with the Effelsberg antenna. Effelsberg observations frequently detected maser emission in particular for o\,Cet (Fig. \ref{fig:ocet-lcurve}) and R\,Leo (Fig. \ref{fig:rleo-lcurve}) at flux density levels below the sensitivity limits of Medicina. Therefore, the maser luminosities $40 \la \log L_{\rm p} [{\rm s}^{-1}] \la 44$ typically observed, are quite likely representing the tip of the iceberg of maser luminosities present in the \water -maser shells of Mira variables and SRVs.

The ratio of the high- and mean-luminosity levels is 3 -- 5 for the Mira variables (U\,Her, RR\,Aql) and the SRVs (R\,Crt, RT\,Vir, SV\,Peg), which did not have observed bursts during our monitoring programme, while this ratio increases to 25 (R\,Cas), $>$8 (R\,Leo) and 6 (RX\,Boo), if bursts occurred.
Excluding the observations made during the burst, the ratio of the high- and mean- luminosity levels for the latter three stars is 3 -- 6, consistent with the ratios obtained for the stars without observed bursts. Thus, the high-luminosity levels are not as representative for the masers as the mean levels, because the former  are significantly increased for stars in which bursts occurred during the monitoring programme. The maser luminosities \Lup\ are also affected by the presence of bursts, as in R\,Leo and R\,Cas a correspondent luminosity increase by a factor 6 -- 7 is noted (see the footnote in Table \ref{table:photon-luminosities}).

We obtained  bolometric fluxes of the stars by SED fitting, as described in \cite{jimenez15}, and determined stellar bolometric luminosities $L_{\rm bol}$ using the distances listed in Table~\ref{table:photon-luminosities}. The mass-loss rates \mdot\  were taken from Table 7 in \cite{debeck10} scaled to the distances used here. These mass-loss rates were determined using the rotationally excited molecular lines of CO and are estimated to have an uncertainty of a factor of 3 (= 0.5 dex) prior to the uncertainties introduced by the distances. 
The results are listed together with those of the SRVs (see Paper~I and II) and the Mira variables U\,Her and RR\,Aql (see Paper~III) in the last two columns of Table~\ref{table:photon-luminosities}. Not surprisingly, there is a tight correlation between \Lup\ and the high values of $L_{\rm p}$. There is also a good positive correlation between $L_{\rm bol}$ and the mass loss rate, and likewise between \Lup\ (and $L_{\rm p}[{\rm High}]$) and \mdot: the larger the mass loss, the stronger the maser luminosity. In view of our earlier conclusion that with our sample we are seeing the tip of the iceberg where maser luminosities are concerned, this latter correlation is more likely to define an upper-envelope, such as found for water-masers in star-forming regions as well \citep{wouterloot95}.

Stellar luminosity and mass-loss rate of $\chi$ Cyg are within the range of the other stars (Table \ref{centralcoords}), so that a reduced abundance of \water\ molecules in the CSE of this S-type star (Sect. \ref{sec:khicyg}) remains the best explanation for the persistent non-detection of \water -maser emission, at least at luminosity levels $\log L_p \ge 40.9$ [s$^{-1}$].

We had found in Paper\,III that the Mira variables U\,Her and RR\,Aql have stellar luminosities, mass-loss rates, and \water -maser photon luminosities within the range shown by the SRVs. This is also true for R\,Cas, although the mean photon luminosity is the lowest within the correspondent range. o\,Cet and R\,Leo are fainter in all three properties, and we only detect them because of their smaller distance to the Sun ($<100$\,kpc) compared to the other stars. 
R\,Leo was selected for the monitoring programme because of its exceptional variability after its discovery (Sect. \ref{sec:rleo}) and o\,Cet because of its outstanding role as leading representative of the Mira-class of long-period variable stars. They represent Mira variables with 10 -- 1000 times fainter \water -maser luminosity compared to the other three Mira variables in our sample. Mira variables like o\,Cet and R\,Leo (or SRVs of similar maser luminosity) but at larger distances, are too faint for regular maser detections with Medicina-class radio telescopes. It's a selection effect that they are under-represented in our monitoring programme.

In part, the selection of stars for monitoring programmes of their \water\-maser emission was likely influenced by the occurrence of bursts during the times of their discovery. As we observed for RX\,Boo, R\,Leo, and R\,Cas, for a couple of months the maser luminosities are enhanced substantially compared to the regular luminosity levels. The cause of their occurrence is not known, but they seem not to be related to regular pulsations, as they are observed in SRV's and Mira variables alike. Furthermore, while the 2010 burst observed in R\,Cas peaked shortly after maximum optical light (Fig. \ref{fig:rcas-logs-main-comps}), the opposite was observed for R Leo, where the burst peaked around minimum optical light (Fig. \ref{fig:rleo-lcurve-two}).

\subsection{\water -maser outflow velocities and velocity shifts}
\begin{figure}
\resizebox{9cm}{!}
{\includegraphics
{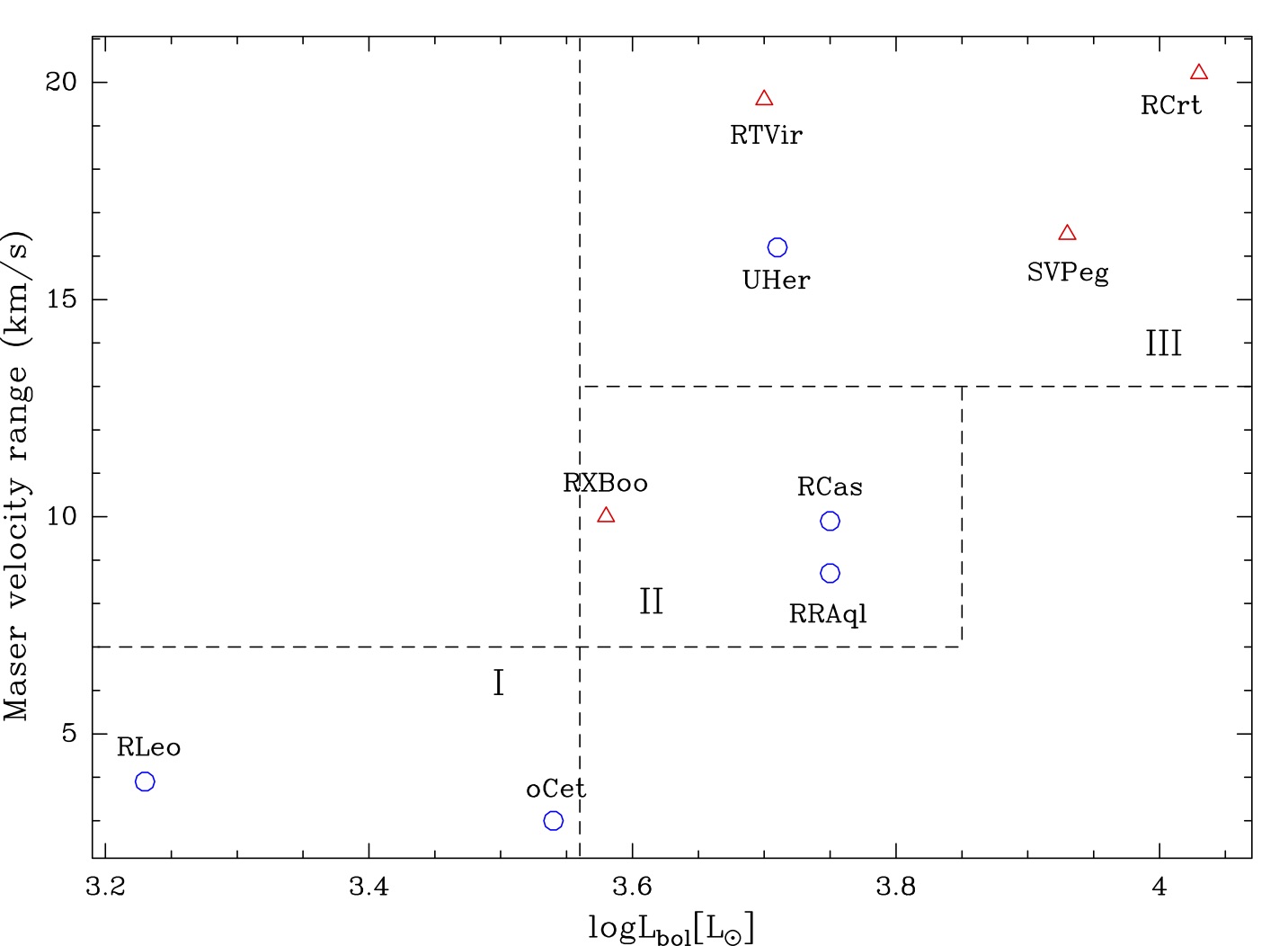}
}
\caption{Velocity range covered by the \water -maser emission vs. bolometric luminosity for the stars discussed in the current paper and in papers I-III. The red triangles and blue circles indicate SRVs and Miras, respectively. The dashed lines mark regions (I-III) for sources whose properties might be different. 
}
\label{fig:lbol-velo}
\end{figure}

The Mira variables that we discuss in this paper show a different behaviour from those appearing in Paper III, U\,Her and RR\,Aql. \\
o\,Cet and R\,Leo have lower $L_{\rm bol}$ and lower \mdot, and thus (see Sect.~\ref{maserlumo}) have a limited capacity for maser excitation. We only see the strongest components, therefore the velocity range of maser emission is narrow. Because the maser emission is not isotropic, we may not see the full \water -maser zone of the CSE. R\,Cas and $\chi$\,Cyg are not particularly weak, but the latter has not been detected at all, while the former shows behaviour not seen before (damped harmonic oscillator). Looking at the bigger picture, plotting the velocity range of the maser emission against $L_{\rm bol}$ (Fig.~\ref{fig:lbol-velo}) we can distinguish three regions occupied by the sources. The clearest difference is found between sources in regions I and III: low-luminosity and narrow emission range (I) versus high luminosity and maser emission over a wider range of velocities (III); region II is a transition between those two extremes. More luminous stars can excite masers at more locations in the CSE and at larger distances from the star, where the outflow velocities are higher, and therefore, we can detect maser emission components over a wider range in velocities. Weaker stars, on the other hand, cannot do this, and we therefore mostly detect the stronger emission from masers that move nearly perpendicular to the line of sight ($V_{\rm los} \approx V_{\ast}$), where the gain path is longest. The detected \water -maser regions in o\,Cet and R\,Leo may thus not trace the full extent of the water-maser shell.

As already introduced in Sect.~\ref{shellkinemat}, we interpret the recurrent damped harmonic oscillator behaviour of the integrated flux density with time of the maser emission in R\,Cas as being the record of  maser clouds moving through the 'favourable excitation zone' in the CSE. The maser emission is dominated by two maser components originating in two clouds,  or cloud groups, that were expelled into opposing hemispheres of the CSE during a mass loss event at some point in time before the start of our monitoring campaigns, and that subsequently moved through the 'favourable excitation zone', leaving it around the year 1997. When we started observing, at the end of 1987, we caught them moving out of the maser zone. During the $\sim$ 8~years that the dominating components were detected, the redshifted component showed a slight decrease in $V_{\rm los}$, which became slightly bluer (by 0.16~\kms). The deviation of the individual velocity components to the fit is very small, suggesting that this concerns a single maser cloud that lived for at least $\sim 8$~years (also in the map of \cite{colomer00} from 1990 there is a single component at this velocity). Likely, a second mass loss event occurred, but this time only the blue component was detected, which could be followed for the next $\sim 20$~years. This component did not show any systematic changes in velocity.

According to the standard model \citep{hoefner18} maser clouds move through the CSE driven by an accelerating stellar wind. 
When monitoring masers in CSEs for a long enough time, one may expect to see systematic velocity shifts in the emission components, if the spectral resolution is good enough. In the stars analysed so far (Table ~\ref{table:photon-luminosities}) systematic velocity drifts were not detected, however. A possible explanation for this, is when clouds have a short lifetime (such that within their lifetime a change in their $V_{\rm los}$ due to their movement in the CSE is of the order of or smaller than the spectral resolution) and they are replaced by other clouds at approximately the same location in the shell and with therefore about the same outflow velocity (see the case of U\,Her discussed in paper III). This is the case when there is a relatively narrow zone in the CSE with favourable maser excitation conditions, or when the acceleration in the water-maser shell is small. Even if a maser cloud has a much longer lifetime, however, it is possible that it does not change its velocity. The ``11~\kms\ component" in RT\,Vir shows (paper II) that for at least 7.5 years a single maser cloud travelled through about half of RT Vir's \water -maser shell without changing its velocity. From this, we inferred that its path was located in the outer part of the water-maser shell, where RT Vir's stellar wind has, apparently, already reached its terminal outflow velocity. In case of the 24.5 \kms\ component of R\,Cas, the more likely cause of the constant $V_{\rm los}$ is its movement nearly perpendicular to the line of sight (the angle $\theta \sim 70\degr - 80\degr$; see Sect.~\ref{shellkinemat}). Judging from the component's small velocity fluctuations,  we are most likely seeing several short-lived maser clouds, which are  parts of a long-living group, expelled into the CSE during a mass loss event, in which a water-maser is excited at different times.

No systematic \water -maser velocity gradients were found either for o~Cet nor for R~Leo. A velocity shift was observed for one of the dominant components in R\,Cas, however. This case is interpreted as the maser cloud moving back towards the star. 
Systematic velocity shifts, were found in the Mira variable IK\,Tau as well, which will be discussed in an upcoming paper (see also \citealt{brand18}). Similar shifts were seen in WX\,Ser \citep{lim24}. Perhaps not surprisingly both these Miras have $L_{\rm bol} > 10^4$~\Lsol, and their maser emission covers a range in velocity of $\sim 31$~\kms\ (IK\,Tau) and $\sim 15$~\kms\ (WX\,Ser), which places them in box III in Fig.~\ref{fig:lbol-velo}. 

\subsection{Limitations of the standard model of the CSE}

We interpreted the \water -maser observations in the context of the standard model of the CSE, which describes the stellar wind as spherically symmetric and assumes a steadily increasing outflow velocity approaching a final value in the outer part of the envelope. While the periodic maser variations do not require modifications to this model, because the maser excitation may respond to the stellar light variations homogeneously throughout the shell, the short-term fluctuations and the long-term (> $P_{\rm opt}$) variations in the maser brightness and the non-systematic variations in the line-of-sight velocities of the maser features demand this. 

The \water -maser emission is fragmented within the \water\ molecular shell within the CSE and the sites of emission might be clouds of enhanced density within the stellar outflow \citep{richards12}, or regions sustaining  favourable excitation conditions on longer (than a few years) timescales, and \water\ molecules are excited while traversing such regions (Paper\,III). ALMA observations of nearby Mira variables (among them o\,Cet and R\,Leo) found complex conditions in the inner parts of the CSE ($\la 3R_*$, i.e. $\sim$5\,au) as well as in the outer envelope at $R>100$ au. Consistently, turbulent motion with outflow and infall and at times velocities exceeding the typical outflow velocities of dust-driven winds is reported for the inner parts of the CSE (o\,Cet: \citealt{wong16}; R\,Leo: \citealt{vlemmings19, hoai23}; as well as in other Mira variables, as for example R\,Dor: \citealt{khouri24}). This happens very likely inside the dust formation radius. The dust driven wind initiates there, accelerates and approaches a final expansion velocity in the outer parts of the CSE. \cite{hoai20} for o\,Cet and \cite{hoai23} for R\,Leo find that the thermal SiO and CO emission in the outer parts is patchy with indications for episodic mass ejections and outflows in particular directions. Such deviations from the spherically symmetric standard model of the CSE are likely also seen in the intermediate \water -maser shell, in particular traced by the long-term brightness variations happening on timescales linked to the crossing times through the \water -maser shell (Paper\,III).

Therefore, monitoring observations of the \water -masers, which probe intermediate radial distances will likely be able to trace the inhomogeneities of the stellar wind, after leaving the turbulent inner part marked by the dust condensation radius. It might become feasible to follow such inhomogeneities in the stellar wind by continuous observations on time scales of many decades corresponding to a length scale of the intermediate region of $\sim$100\,au and outflow velocities $5-10$ \kms. 

\section{Summary}

We have analysed the water-maser emission in three Mira variables. In a fourth Mira ($\chi$\,Cyg), no water-maser has (ever) been detected. Compared to the Miras discussed in paper III (U\,Her and RR\,Aql), the properties are different in terms of detection rates, variability, and velocity range of the maser emission.

Where we were able to determine a maser period, the maser emission follows the stellar pulsations with the same period, but with a phase lag of $0.32 - 0.33$. The \water -maser emission in R\,Leo was below the (Medicina) sensitivity limits for fairly long periods of time, which prevented us from determining a radio period. The maser of R\,Cas also had relatively long time intervals in which it was below our sensitivity levels, but when it was active, it had a well-determined periodic brightness variation.
In addition to the periodic variations in flux, infrequent bursts may occur (with a duration of about one year): At least three maser burst episodes occurred during 3.5 decades of observations in R Cas; one burst occurred in R\,Leo in about 2.5 decades, and no bursts were seen in o\,Cet in three decades. $\chi$\,Cyg was not detected at all.

R\,Cas is a special case among the stars in our sample: While the maser emission follows the pulsations of the star, from the start of our observations its integrated flux density became progressively weaker, showing the characteristics of a damped harmonic oscillator. After a subsequent relatively brief period of low activity, the total flux increased again, following the same pattern in reverse. The emission reached a maximum and then declined again. This behaviour has not been seen before. We interpret it as cloud(s) moving through a zone of favourable maser excitation conditions, the existence of which was discussed in paper III. We estimate that the outer edge of this zone is located at $\sim$20\,au. 

Of the stars we discussed, R Cas alone has a maser component showing a (very small) systematic velocity gradient. We identify this component as originating in a single cloud that falls back towards the star, with a lifetime of at least about eight years.

The maser emission in the three stars we studied has a fairly narrow velocity range, although the emission in R\,Cas is somewhat wider in the late 1980s to early 1990s. This is true in particular for the stars with a low bolometric luminosity, o\,Cet and R\,Leo. The bolometric luminosity of a star and the velocity range of the water-maser emission are related: More luminous stars excite more maser locations in the shell (and at greater distances from the star). We see only the strongest components in R Leo and o Cet, and thus, only part of the water-maser shell is visible to us.

\section{Data availability}
The maser spectra are available at the CDS via anonymous ftp to cdsarc.u-strasbg.fr (130.79.128.5) or via http://cdsweb.u-strasbg.fr/cgi-bin/qcat?/J/A+A/. The Appendix with all spectra is available on \href{https://zenodo.org/records/15534987}{https://zenodo.org/records/15534987}.

\begin{acknowledgements}
The Medicina 32 m data presented here pertaining to 1987-2011, were obtained during a long-term monitoring programme, which concerned both late-type stars and star-forming regions.
Thanks to those who helped with the observations. We are grateful to the staff at the Medicina observatory for their expert assistance and technical problem-solving. The Medicina "Grueff" radio telescope is funded by the Ministry of University and Research (MUR) and is operated as National Facility by the National Institute for Astrophysics (INAF).
This research is partly based on observations with the 100 m telescope of the Max-Planck-Institut für Radioastronomie (MPIfR) at Effelsberg, and has made use of the SIMBAD database, operated at CDS, Strasbourg, France. This research has made use of NASA's Astrophysics Data System. 
For data reduction and the preparation of figures GILDAS software, available at www.iram.fr/IRAMFR/GILDAS, was used. We acknowledge with gratitude the variable star observations from the AAVSO International Database, contributed by observers worldwide and used in this research. This document was prepared with the web application Overleaf [www.overleaf.com]. 
\end{acknowledgements}

\bibliographystyle{aa}
\bibliography{masterlist-biblio.bib}

\end{document}